\documentclass[preprintnumbers,amsmath,amssymb,aps,prd]{revtex4}
\pdfoutput=1
\usepackage{graphicx}
\usepackage{amsmath,amssymb}
\usepackage{epstopdf}
\usepackage{color}%

\usepackage[colorlinks,urlcolor=blue,citecolor=blue]{hyperref}
\def\hhref#1{\href{http://arxiv.org/abs/#1}{arXiv:#1}} 

\def\mc{\mathcal}

\begin{document}

\title{WKB and Resurgence in the Mathieu Equation}

\author{Gerald V. Dunne}
\affiliation{Department  of Physics, University  of Connecticut, Storrs, CT, 06269}
\author{Mithat \"Unsal}
\affiliation{Department  of Physics, North Carolina State University, Raleigh, NC, 27695\\
 Department of Mathematics, Harvard University, Cambridge, MA, 02138
}


\begin{abstract}
In this paper, based on lectures by the authors at the May 2015 workshop {\it Resurgence, Physics and Numbers}, at the Centro di Ricerca Matematica Ennio De Giorgio of the Scuola Normale Superiore in Pisa,  we explain the origin of resurgent trans-series in the Mathieu equation spectral problem, using uniform WKB and all-orders (exact) WKB. 
Exact quantization conditions naturally arise, and their expansion in the weak coupling regime produces resurgent trans-series expressions which exhibit precise relations between different instanton sectors. Indeed,  the perturbative expansion encodes all orders of the multi-instanton expansion, an explicit realization of the general concept of ``resurgence''.  We also discuss the transition from weak to strong coupling, an explicit realization of  ``instanton condensation''.

\end{abstract}


\maketitle

\section{Introduction}

\subsection{Some Motivation from Physics and Mathematics}

Asymptotic analysis is a cornerstone of physics, providing accurate expansions when certain dimensionless combinations of physical parameters are large or small. Such expansions are often divergent, and these divergent expansions contain a wealth of information, in addition to giving numerically accurate estimates in the appropriate limiting regimes. The modern mathematical theory of resurgent asymptotics, based on Borel-Ecalle summation, is based on trans-series expansions, in which formal (divergent) series expansions in a small parameter are extended to trans-series expansions that also include summations over exponentially suppressed non-perturbative terms, as well as possible summations over powers of logarithms \cite{Ecalle:1981,s07,Costin:2009,delabaere,Sternin:1996}. While this mathematical theory arose from quite abstract origins, it now appears that it is surprisingly well suited to many problems in physics. For example, in the language of quantum field theory,  {\it multi-instanton calculus} is a trans-series expansion, involving a sum over an infinite set of non-perturbative multi-instanton sectors, each of which is multiplied by a fluctuation series, and some of which are multiplied by series of logarithms, due to quasi-zero modes. In another context, it is well known in physics that summation of classes of Feynman diagrams often leads to series in powers of logarithms. From the physics perspective, the primary new idea from the formal theory of resurgence is the notion that there should generically be precise relationships between the fluctuations about different non-perturbative sectors, including the vacuum sector (``perturbation theory''). Viewed from the path integral, this is quite unexpected. In these lectures we explain in simple terms how this type of structure arises, using the classic example of instantons in the quantum spectral problem for the periodic cosine potential, known in mathematics as the Mathieu equation \cite{ww,goldstein,mclachlan,meixner,magnus,nist-mathieu}. Using various forms of WKB, extended to all orders, we show how the trans-series structure arises. In the process, we find that not only is there a natural underlying trans-series, but there are concrete quatitative relations between different instanton sectors. Indeed, in an extreme manifestation of resurgence, we show that all orders of the trans-series are encoded in a subtle way in the divergent perturbative expansion itself. Here we explain what we mean by this somewhat dramatic claim. We use the language of the underlying differential equation, but these conclusions are even more interesting when translated into the path integral formalism. We propose that these ideas could in fact be used as a means to provide a sensible mathematical definition of the path integral, even in Minkowski space, and moreover a definition that provides also a means of making quantitative computations. One of the primary motivations is that in many physics applications it would be desirable to have better control over the analytic continuation of path integrals, for example for studying real-time processes, quantum transport, or finite density systems. One of the main mathematical advantages of a trans-series, compared to just an ordinary divergent series, is that the trans-series is designed to encode the proper analytic continuation properties of the function it is describing. This sounds appealing, but it is a notoriously difficult problem in path integrals, so we are motivated to understand as much of this trans-series structure as possible, in a well-defined  concrete example. That is the purpose of these lectures.

The study of resurgent trans-series in quantum spectral problems began with the pioneering work of Bogomolny and Fateev, and Zinn-Justin, Br\'ezin, Parisi, Voros et al \cite{Balian:1978et,Bogomolny:1980ur,ZinnJustin:1981dx,voros,Balitsky:1985in,zinnbook} (for an excellent review see \cite{LeGuillou:1990nq}), as well as in the mathematical literature from the work of Pham, Ramis, Dillinger, Delabaere, Berry, Howls, Aoki, Takei et al \cite{BerryHowls,delabaere-periodic,ddp,ramis,Aoki:1993ra,howls-book}. It also became clear from the work of Kruskal and Costin and co-workers, that resurgent asymptotic analysis is a powerful tool in studying the asymptotics of nonlinear differential equations \cite{costin-nl,Costin:2009,howls-book}.
More recently, in physics, Mari\~no, Schiappa and Weiss showed that the resurgent approach yields interesting new insights in the study of matrix models and string theory \cite{Marino:2007te,Marino:2008vx}. Since then there has been a flourishing activity involving applications in matrix models, string theory, quantum gauge theory and sigma models, as well as new results in quantum mechanics \cite{Pasquetti:2009jg,Marino:2012zq,Aniceto:2011nu,Argyres:2012ka,Dunne:2012zk,Cherman:2013yfa,Misumi:2014jua,Aniceto:2014hoa,Couso-Santamaria:2014iia,Couso-Santamaria:2015wga,Basar:2015xna,on}. In these lectures, our goal is quite modest: study a well-known and widely relevant system, the Mathieu equation, and understand it in great detail. Surprisingly, through the eyes of resurgence we are able to see new results in this very old problem.

Two further pieces of motivation are the following general questions:
\begin{itemize}
\item
What can  weak coupling analysis tell us about strong coupling ?

\item Can weak coupling and  strong coupling be related, even if the degrees of freedom re-organize themselves in a very non-trivial way, for example across a transition ?
\end{itemize}
We will be able to phrase these questions more precisely in the context of the Mathieu system, and see that resurgent asymptotic analysis provides precise answers.

\subsection{Resurgent trans-series in quantum spectral problems}

In this paper we concentrate on trans-series expressions for energy eigenvalues in 
certain quantum mechanical spectral problems, more specifically the Schr\"odinger equation for the periodic cosine (Mathieu) potential. 
\begin{eqnarray}
 -\frac{\hbar^2}{2} \frac{d^2}{dx^2}\psi(x)+\cos(x) \psi(x)=u\, \psi(x)
\label{eq:mathieu}
\end{eqnarray}
This is an example of a more general class of quantum mechanical (QM) spectral problems with potentials having degenerate harmonic minima. In these cases it is known that standard Rayleigh-Schr\"odinger perturbation theory leads to a perturbative series that is not only divergent, but also Borel non-summable, in the sense that the expansion coefficients grow factorially fast in magnitude and do not alternate in sign  \cite{Brezin:1977gk,Stone:1977au,Bogomolny:1980ur,ZinnJustin:1981dx,ZinnJustin:2004ib,Unsal:2012zj}. Thus, naive Borel summation leads to a non-perturbative imaginary part for the energy, which moreover is ambiguous. This is doubly problematic: not only is it ambiguous, but these systems are stable, and therefore the energy should be real. (Contrast with the case of a cubic oscillator \cite{arkady}, or an inverted-double-well quartic oscillator \cite{bw}, where the system is unstable and the imaginary part has a natural physical interpretation.) The resurgent trans-series expression for the energy resolves these apparent inconsistencies, because the perturbative series is only one part of the full trans-series, and the ambiguous imaginary non-perturbative term arising from Borel summation of perturbation theory is exactly cancelled by a corresponding term in a higher non-perturbative sector. Resurgence implies that these cancellations occur to all orders, leading to a full trans-series that is not only real, but also unambiguous.  This means that perturbation theory on its own is incomplete, while the full trans-series is complete. For a clear and  exhaustive analysis of how these cancellations occur for general real trans-series we refer the reader to \cite{Aniceto:2013fka}.

In the context of the Mathieu equation, consider performing perturbation theory within one of the potential wells, perturbing about the $N^{\rm th}$ energy level of the unperturbed harmonic well, leading to a perturbative expression of  the form
\begin{eqnarray}
u_{\rm pert}(\hbar, N)=\sum_{n=0}^\infty \hbar^{n} u_{n}(N)
\label{eq:perturbation}
\end{eqnarray}
Notice that the system can be scaled in such a way that the sole parameter in the equation is $\hbar$, and the ``semiclassical limit'' of small $\hbar$ refers to the situation where the potential wells are deep. (See Section \ref{sec:notation} below.)
The perturbative coefficients $u_{n}(N)$ are simple polynomials in the level number $N$, and can be computed by straightforward iterative procedures. For potentials with degenerate harmonic minima, the perturbative expansion (\ref{eq:perturbation}) is not Borel summable, which means that on its own it is incomplete and indeed inconsistent. This is true not just for the ground state ($N=0$), but also for higher states, so long as $N\ll 1/\hbar$, as discussed in more detail in Section 
\ref{sec:wkb} below. This situation can be remedied by recognizing that the full expansion of the energy at small coupling is in fact of the ``trans-series'' form:
\begin{eqnarray}
u_{\rm trans}(\hbar, N)=
\sum_{k=0}^\infty \sum_{n=0}^\infty \sum_{l=1}^{k-1} c_{k, n,  l}(N) \, \hbar^n \left(\frac{1}{\hbar^{N+1/2}}\,\exp\left[-\frac{S}{\hbar}\right]\right)^k 
\left(\ln \left[-\frac{1}{\hbar}\right]\right)^l
\label{trans}
\end{eqnarray}
Perturbation theory corresponds to the ``zero-instanton sector'', $k=0$, with coefficients $c_{0, n, 0}(N)\equiv u_n(N)$.
The higher ($k\geq 1$) instanton terms of the trans-series involve a sum over non-perturbative factors $\exp[-k\, S/\hbar]$, multiplied by prefactors that are themselves series in $\hbar$ and in $\ln(1/\hbar)$. Note that the sum over logarithms in (\ref{trans}) begins in the $k=2$ sector: physically, these logarithms arise from the interaction between instantons and anti-instantons, which requires $k\geq 2$.

The basic building blocks of the trans-series,  $\hbar$, $\exp[-S/\hbar]$  and $\ln(1/\hbar)$, are called ``trans-monomials'', and are  familiar from physical examples.  The parameter $S$ in (\ref{trans}) is a numerical constant, the single-instanton action. With our choice of scaling of the Mathieu equation, $S=8$ [see equation (\ref{eq:s}) below].
Remarkably, the expansion coefficients $c_{k, n, l}(N)$ of the trans-series are intertwined amongst themselves, in such a way that the `necessary' cancellations occur in order to render the full  trans-series real and unambiguous. In practice, this works as follows: a Borel analysis of the perturbative series requires an analytic continuation in $\hbar$, producing non-perturbative imaginary parts, but these are precisely cancelled by the imaginary parts associated with the $\ln(-1/\hbar)$ factors in the non-perturbative portion of the trans-series. 
Similarly, the fluctuations about the single instanton sector, given by the coefficients $c_{1, n, 0}$, are also divergent, and Borel summation of these fluctuations produces new imaginary parts, but these are cancelled by terms in the $k=3$ sector. And so on. Ambiguities only arise if you look at just one isolated portion of the trans-series expansion, for example just the perturbative part, or just some particular multi-instanton sector. When viewed as a whole, the analytic continuation of the trans-series expression is real, unique and exact.

An important  first step in our argument  is a seemingly small and innocent shift of emphasis from much of the previous work studying the divergence of perturbation theory in quantum spectral problems, which has often concentrated on low-lying energy levels or bands, such as the ground state or lowest band. In order to see the full structure of the trans-series it proves useful to consider the spectral energy eigenvalue not just as a function of the small coupling (which we can take here as $\hbar$: see the scaling defined in Section \ref{sec:notation} below), but also of the level or band number, $N$, an integer that labels the perturbative energy level or band. The fact that there exists such a label is physically clear for problems with degenerate harmonic minima, corresponding just to the unperturbed harmonic oscillator level number. Mathematically, it is clear from oscillation theorems for Sturm-Liouville type problems in one dimension. Thus we view the energy eigenvalue $u$ in (\ref{eq:mathieu}) as a function of two variables:
\begin{eqnarray}
u=u(\hbar, N)
\label{eq:unh}
\end{eqnarray}
This immediately defines three interesting, and quite distinct, spectral regions:
\begin{enumerate}
\item
$N \hbar\ll 1$: weak coupling, far below the barrier
\item
$N \hbar \sim O(1)$: intermediate coupling, near the barrier top
\item
$N \hbar \gg 1$: strong coupling, far above the barrier
\end{enumerate}
Note that even the physical language used to describe the different spectral regions is very different. In the weak coupling region the states are described as superpositions of localized atomic states in the so-called ``tight-binding approximation'', while at strong-coupling the states are more naturally treated as using the ``nearly-free electron'' picture of weakly perturbed free states \cite{kittel}. Near the barrier top, neither of these approximations is satisfactory, and there is a complicated re-arrangement of degrees of freedom, a concrete realization of the phenomenon of  ``instanton condensation''  \cite{gw,wadia,neuberger,matytsin,witten}. Correspondingly, the expansions of the energy in these three regions are quite different, and yet they are all related. One of the main motivations of this work is to understand in more precise detail how this transition from weak- to strong-coupling occurs. In particular, we are motivated by the close analogy to so-called ``large $N$'' technqiues in quantum gauge theories and matrix models \cite{marcos-book}.

So far, the discussion sounds somewhat trivial. But the new result that we wish to describe is that by considering the dependence of $u(\hbar, N)$ on {\it both} $\hbar$ and $N$ we find that there is a direct quantitative relation between perturbation theory and the all-orders multi-instanton trans-series expression. For example, one explicit consequence of this relation is a dramatically improved description of the one-instanton sector. 
\\

\begin{quote}
{\it New Result for Mathieu Spectrum} \cite{Dunne:2014bca,Dunne:2013ada}: The leading exponential (``one-instanton'') splitting of the $N^{\rm th}$ band in the weak-coupling regime where $N \hbar\ll 1$ can be written, including the series of fluctuations,
\begin{eqnarray}
u(\hbar, N)=u_{\rm pert}(\hbar, N) \pm   \frac{\hbar}{\sqrt{2 \pi}} \frac{1}{N!} \left(\frac{32}{\hbar}\right)^{N+\frac{1}{2}} \exp\left[-\frac{S}{\hbar}\right]\,{\mathcal P}_{\rm inst}(\hbar, N)\,  +\dots
\label{eq:one}
\end{eqnarray}
where  the fluctuation factor ${\mathcal P}_{\rm inst}(\hbar, N)$ is expressed entirely in terms of the perturbative expansion $u_{\rm pert}(\hbar, N)$:
\begin{eqnarray}
{\mathcal P}_{\rm inst}(\hbar, N)=\frac{\partial u_{\rm pert}(\hbar, N)}{\partial N}\,  \exp\left[S\int_0^{\hbar} \frac{d\hbar}{\hbar^3}\left(\frac{\partial u_{\rm pert}(\hbar, N)}{\partial N}-\hbar+\frac{\left(N+\frac{1}{2}\right)\, \hbar^2}{S}\right)\right] \, , \quad S\equiv S_{\rm instanton}=8
\label{eq:prefactor}
\end{eqnarray}
\end{quote}
This result (\ref{eq:one}, \ref{eq:prefactor}) agrees with the fluctuation series derived from the asymptotics of Mathieu functions \cite{muller}, and also with a recent 3-loop computation for $N=0$ \cite{Escobar-Ruiz:2015rfa}, as discussed below in Section \ref{sec:confirm}. The interesting new thing in (\ref{eq:one}, \ref{eq:prefactor}) is that the fluctuation series ${\mathcal P}_{\rm inst}(\hbar, N)$ is expressed solely in terms of the perturbative fluctuation series $u_{\rm pert}(\hbar, N)$.
Thus, there is a direct relation between the fluctuations about the zero-instanton sector (i.e., perturbation theory, $u_{\rm pert}(\hbar, N)$), and the fluctuations about the one-instanton sector, ${\mathcal P}_{\rm inst}(\hbar, N)$. This is an explicit example of resurgence, quite different from the early term/late term relations that have been studied previously.

The first factor in (\ref{eq:prefactor}), $\partial u_{\rm pert}/\partial N$, is a density-of-states factor that is well known at leading order in $\hbar$ \cite{keller,connor}, but the remaining exponential factor in (\ref{eq:prefactor}) is new \cite{Dunne:2014bca,Dunne:2013ada}. 
Similar relations can be derived expressing the fluctuations in any higher-instanton sector exactly in terms of the fluctuations about the zero-instanton sector (i.e., perturbation theory). Furthermore, there are similar results for other potentials \cite{Dunne:2014bca,Dunne:2013ada}. This is remarkable: it says that perturbation theory, $u(\hbar, N)$, encodes everything! One simply has to know how to decode this information.

In Section \ref{sec:basic} we review basic known facts about the Mathieu spectrum. In Sections \ref{sec:uniform} - \ref{sec:wkb}  we illustrate and derive the results (\ref{eq:one}, \ref{eq:prefactor}) in the context of the Schr\"odinger equation for these quantum mechanical systems, but the result is even more interesting when interpreted in the language of the path integral approach to the same spectral problem, as discussed in Section \ref{sec:path}. This is in fact our main motivation, as we wish to develop a deeper insight into the general structure of semiclassical expansions, having in mind potential applications to quantum field theory.

\section{Basic facts about the Mathieu spectrum}
\label{sec:basic}

Here we summarize the basic classic results concerning the spectrum of the Mathieu equation (see e.g.  \cite{ww,goldstein,mclachlan,meixner,magnus,muller,nist-mathieu}).

\subsection{Notation and Scaling}
\label{sec:notation}

\begin{figure}[htb]
\centering
\includegraphics[scale=.7]{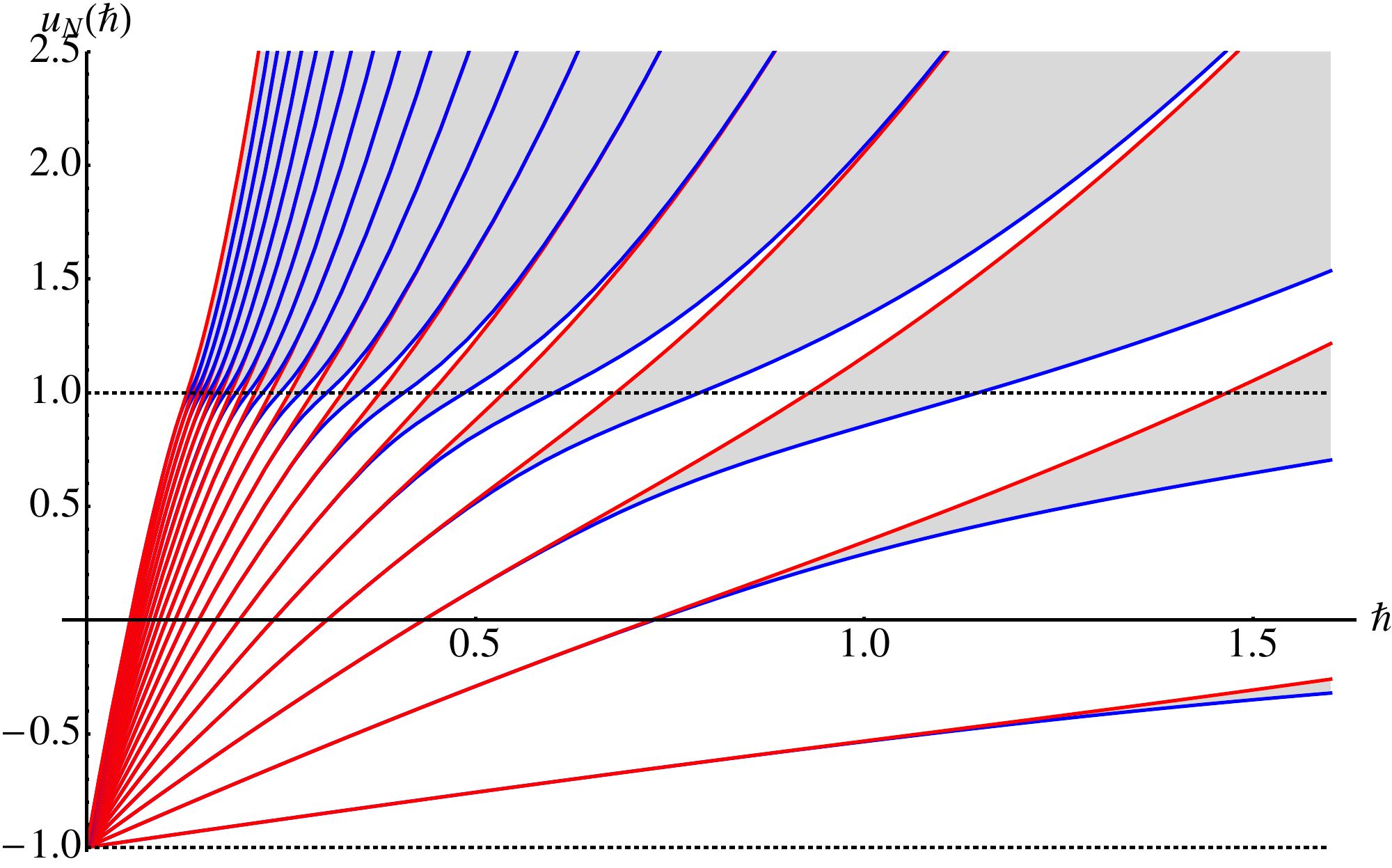}
\caption{The energy spectrum of the Mathieu equation,  as a function of the parameter $\hbar$. The {\it regions of stability} (the bands) are shaded, with  lower edges shown as solid (blue) lines, and top edges shown  as  solid (red) lines. The bands are separated by {\it regions of instability} (gaps), which are unshaded. The first 20 bands are shown. At small $\hbar$, the bands are exponentially narrow, and the band {\it location} follows the linear behavior in (\ref{eq:weak}). High in the spectrum, the gaps are exponentially narrow, and the gap {\it location} follows the quadratic behavior in (\ref{eq:strong}). 
The maximum and minimum of the potential, at $u=\pm 1$, are shown as dotted lines.
Note the smooth transition, near $u\sim 1$, between narrow bands at  small $\hbar$, and  narrow gaps  at large $\hbar$. In this region, the bands and gaps are of equal width, and are not exponentially narrow, as discussed Section \ref{sec:top}.
}
\label{fig:fig1}
\end{figure}

The Mathieu equation describes the nonlinear oscillator problem, and has a wide array of applications \cite{meixner,nist-mathieu}.  We write the Mathieu equation in the Schr\"odinger  equation form (\ref{eq:mathieu}), but by simple changes of variables,  this becomes the standard textbook form of the Mathieu equation \cite{nist-mathieu}:
\begin{eqnarray}
\frac{d^2\psi}{dz^2}+\left(A-2Q\, \cos(2z)\right)\psi=0\quad\longleftrightarrow\quad -\frac{\hbar^2}{2}\,\frac{d^2\psi}{dx^2}+ \cos(x)\psi=u\, \psi
\label{mathieu1}
\end{eqnarray}
(We use capital letters $A$ and $Q$, rather than  the conventional lower-case ones, to avoid confusion with symbols $a$ and $q$, which have special meaning in the related  gauge theory discussion).
So we can translate back and forth between notations with  the  identifications:
\begin{eqnarray}
Q=\frac{4}{\hbar^2}\qquad, \qquad A=\frac{8 \,u}{\hbar^2}
\label{hbar}
\end{eqnarray}
We also wish to compare with the important work of Zinn-Justin and Jentschura \cite{ZinnJustin:2004ib},
who used yet another scaling:
\begin{eqnarray}
\left(-\frac{1}{2}\frac{d^2}{dx^2}+\frac{1}{8 g}\sin^2(2 \sqrt{g}\, x)\right)\psi= E_{ZJJ}\, \psi
\quad\longleftrightarrow\quad 
\left(-\frac{(16g)^2}{2}\frac{d^2}{dx^2}+\cos(x)\right)\psi=(16g\, E_{ZJJ}-1)\, \psi 
\label{zj}
\end{eqnarray}
Thus  $\hbar$ plays the role of a coupling constant $g$:
\begin{eqnarray}
\hbar =16\, g \qquad, \qquad u=-1+16\,g\, E_{ZJJ} =-1+\hbar \, E_{ZJJ}
\label{hbar-g}
\end{eqnarray}
The Mathieu system has a QM spectrum consisting of an infinite series of bands and gaps (also known as regions of stability and instability), as shown in Figure \ref{fig:fig1}. Low in the spectrum the bands are very narrow, and high in the spectrum the gaps are very narrow. The transition between these two behaviors occurs near the top of the potential barrier, at $u=1$, where the bands and gaps are of approximately equal width. These features are explained quantitatively below, in Section \ref{sec:wkb}. We are particularly interested in the transition between the two extreme regions, which occurs near $u=1$, the maximum of the potential $\cos x$.

\subsection{Weak Coupling: {$N\hbar\ll 1$}}
\label{sec:weak}

At small $\hbar$, with $N\hbar\ll 1$, we effectively have far-separated harmonic wells, with high barriers between them. Then perturbation theory in a  given well, starting with the $N^{\rm th}$ harmonic oscillator level, leads to the following perturbative expression \cite{goldstein,meixner,nist-mathieu,muller,mclachlan}:
\begin{eqnarray}
 u_{\rm pert}(\hbar, N) 
&\sim& -1 +\hbar \left[N+\frac{1}{2}\right] -\frac{\hbar^2}{16}\left[\left(N+\frac{1}{2}\right)^2+\frac{1}{4}\right]-\frac{\hbar^3}{16^2}\left[\left(N+\frac{1}{2}\right)^3+\frac{3}{4}\left(N+\frac{1}{2}\right)\right] \nonumber\\
&&  - \frac{\hbar^4}{16^3}\left[ \frac{5}{2}\left(N+\frac{1}{2}\right)^4+\frac{17}{4}\left(N+\frac{1}{2}\right)^2+\frac{9}{32}\right]  \nonumber\\
&& 
-  \frac{\hbar^5}{16^4}\left[ \frac{33}{4}\left(N+\frac{1}{2}\right)^5+\frac{205}{8}\left(N+\frac{1}{2}\right)^3+\frac{405}{64}\left(N+\frac{1}{2}\right)\right] -\dots
\label{eq:weak}
\end{eqnarray}
This expansion is indeed of the form of the perturbative expansion (\ref{eq:perturbation}), 
where at $n^{\rm th}$ order of perturbation theory, the expansion coefficient $u_n(N)$ is a polynomial of degree $n$ in the level number $N$. There is an implicit assumption here that $N\hbar\ll 1$.

For fixed $N$, the $u_n(N)$ are non-alternating in sign and diverge factorially fast \cite{Dunne:2014bca,Dunne:2013ada,muller,Stone:1977au}:
\begin{eqnarray}
u_n(N)&\sim& -\frac{2^{2N}}{\pi \left(N!\right)^2}\frac{\Gamma(n+2N+1)}{16^{n+2N+1}}
\qquad, \quad n\to \infty
\label{eq:mathieu-growth}
 \end{eqnarray}
 This means that the perturbative expansion is Borel non-summable, and so the perturbative expression (\ref{eq:weak}) must be extended to a  trans-series.
The  rate of divergence (\ref{eq:mathieu-growth}) of perturbation theory has a characteristic form, with the factor $16$ being equal to twice the instanton action, where the instanton action for the Mathieu potential is: 
\begin{eqnarray}
S\equiv S_{\rm instanton}=\sqrt{2}\int_{-\pi}^\pi dx\, \sqrt{\cos(x)+1} =8
\label{eq:s}
\end{eqnarray}
In this spectral regime, well below the barrier top, the spectrum consists of narrow bands, whose central location is given by (\ref{eq:weak}), and whose widths  are given by the  classic result \cite{goldstein,nist-mathieu,meixner,muller}
  \begin{eqnarray}
\Delta u_{\rm band}( \hbar, N)
 \sim  \frac{2 \hbar}{\sqrt{2 \pi}} \frac{1}{N!} \left(\frac{32}{\hbar}\right)^{N+1/2}
\exp\left[-\frac{8}{\hbar}\right]\left\{ 1-\frac{\hbar}{32}\left[3 \left(N+\frac{1}{2}\right)^2+4 \left(N+\frac{1}{2}\right) +\frac{3}{4}\right]+O(\hbar^2) \right\}
\label{eq:band-width}
\end{eqnarray}
Physically, we interpret this as a non-perturbative single instanton term \cite{coleman,Neuberger:1978ft}, incuding the fluctuations about the single-instanton. This {\it single-instanton} effect is real and unambiguous. The factor $8$ in the exponent is the instanton action for the Mathieu potential in (\ref{eq:s}).

\subsection{Strong Coupling: $N\hbar\gg 1$}

 In the strong coupling regime, $N\hbar\gg 1$,  far above the barrier top,  the spectral behavior is completely different, as can be seen immediately from Figure \ref{fig:fig1}. 
 We can probe this region by  taking $\hbar$ large, keeping $N$ small and fixed,  in which case the gap edges are (see \cite{nist-mathieu}, converted to our notation):
 \begin{eqnarray}
u_0&=&{\hbar^2\over8}\left(0-\frac{1}{\hbar^2}+\frac{7}{4 \hbar ^6}-\frac{58}{9 \hbar ^{10}}+ \frac{68687}{2304 \hbar ^{14}}+\dots\right)\nonumber \\ 
u^{(-)}_1&=&{\hbar^2\over8}\left(1 - {4\over \hbar^2}- {2\over\hbar^4} + {1\over\hbar^6} - {1\over 6\hbar^8} - {11\over
 36 \hbar^{10}} + {49\over144 \hbar^{12}} - {55\over576 \hbar^{14}} - {83\over 540 \hbar^{16}}+\dots\right)\nonumber \\ 
u^{(+)}_1&=&{\hbar^2\over8}\left(1 + {4\over \hbar^2} - {2\over \hbar^4} - {1\over\hbar^6} - {1\over6 \hbar^8} + {11\over
 36 \hbar^{10}} + {49\over 144 \hbar^{12}} + {55\over576 \hbar^{14}} - {83\over 540 \hbar^{16}}+\dots \right)\nonumber \\ 
u^{(-)}_2&=&{\hbar^2\over8}\left(4 - {4\over 3 \hbar^4} + {5\over 54 \hbar^8} - {289\over 19440 \hbar^{12}} + {21391\over 6998400 \hbar^{16}}+\dots\right)\nonumber \\ 
u^{(+)}_2&=&{\hbar^2\over8}\left(4 + {20\over3 \hbar^4} - {763\over54 \hbar^8} + {1002401\over
 19440 \hbar^{12}} - {1669068401\over6998400 \hbar^{16}}+\dots\right)\nonumber \\ 
u^{(-)}_3&=&{\hbar^2\over8}\left(9 + {1\over\hbar^4} - {1\over\hbar^6} + {13\over80 \hbar^8} + {5\over
 16 \hbar^{10}} - {1961\over5760 \hbar^{12}} +{ 609\over6400 \hbar^{14}}+\dots\right)\nonumber \\ 
u^{(+)}_3&=&{\hbar^2\over8}\left(9 + {1\over\hbar^4} + {1\over\hbar^6} + {13\over80 \hbar^8} - {5\over
 16 \hbar^{10}} - {1961\over5760 \hbar^{12}} - {609\over6400 \hbar^{14}}+\dots\right)\nonumber \\ 
 u^{(-)}_4&=&{\hbar^2\over8}\left( 16 +{ 8\over15 \hbar^4} - {317\over3375 \hbar^8} + {80392\over
 5315625 \hbar^{12}}+\dots\right)\nonumber\\
 u^{(+)}_4&=&{\hbar^2\over8}\left( 16 +{ 8\over15 \hbar^4} +{433\over3375 \hbar^8} - {45608\over5315625 \hbar^{12}}+\dots\right)
\label{eq:conv}
\end{eqnarray}
This is the regime where the kinetic energy dominates the potential energy, so the energy eigenvalues are obtained by perturbing around the (degenerate) free-particle-on-a-circle states, 
with the potential $\frac{2}{\hbar^2} \cos(x)$ treated as a perturbation.  The expansions (\ref{eq:conv}) are {\it convergent}, with a radius of convergence that increases quadratically with the level index $N$. They are conventionally expressed as continued-fraction representations of the eigenvalues, generated by a  Fourier decompostion of the gap-edge eigenfunctions, and these continued-fraction expressions are themselves convergent \cite{ww,nist-mathieu,mclachlan,meixner}.
Nevertheless, despite these convergence properties, there are also non-perturbative effects, associated with the narrow splittings of the spectral gaps in this spectral region, as seen in Figure \ref{fig:fig1}.

Rather than taking $\hbar\gg1$ with $N$ fixed, the high spectral region could also be probed by taking $\hbar\to 0$ and $N\to\infty$, but with $N\hbar\gg 1$ fixed. This is motivated by analogous ``large-$N$'' expansions in quantum gauge theories and matrix models \cite{marcos-book}. We will see that this is a surprisingly close analogy to the Matheiu spectrum.
In this limit, for large level number $N\gg 1/\hbar$,  the continued-fraction expressions for the energy eigenvalues give  approximate expressions for the energy of the $N^{\rm th}$  gap as \cite{nist-mathieu}:
 \begin{eqnarray}
u( \hbar, N) 
& \sim &
\frac{\hbar^2}{8}\left(N^2+\frac{1}{2(N^2-1)}\left(\frac{2}{\hbar}\right)^4
+\frac{5N^2+7}{32(N^2-1)^3(N^2-4)} \left(\frac{2}{\hbar}\right)^8 
\right. \nonumber\\&&\left .
\qquad +\frac{9N^4+58 N^2+29}{64(N^2-1)^5(N^2-4)(N^2-9)}\left(\frac{2}{\hbar}\right)^{12}+\dots\right)
\label{eq:strong}
\end{eqnarray} 
This is an expansion in powers of $(2/\hbar)^4$.
We observe  that the $n^{\rm th}$  coefficient has poles at  certain integer values of $N\leq n$, so the expression (\ref{eq:strong}) should really be interpreted as an expansion about $N=\infty$.
We discuss this further in Section \ref{sec:summary}. We also note that similar expansions, with coefficients with analogous poles, occur in gauge theory large-$N$ expansions \cite{deWit:1977hx,goldschmidt,Samuel:1980vk}.
  
The strong-coupling expansion (\ref{eq:strong}) does not  fully describe  the spectrum. It does not distinguish between the  gap edges, $u_N^{(\pm)}$. Furthermore, the $\hbar$ dependence of the expansion (\ref{eq:strong}) is different from that in (\ref{eq:conv}). 
For a given $N$, there is a non-perturbative splitting of the gap, occuring  at  order $1/\hbar^{2N}$. Thus, as the level index $N$ increases, the gap splitting occurs at higher orders in perturbation theory, becoming exponentially small \cite{nist-mathieu}:
\begin{eqnarray}
\Delta u_{\rm gap}(\hbar, N)
&\sim& \frac{\hbar^2}{4} \frac{1}{\left(2^{N-1} (N-1)! \right)^2} \left(\frac{2}{\hbar}\right)^{2N}  \left[1+O\left(\left(\frac{2}{\hbar}\right)^4\right) \right]  \nonumber \\
&\sim & \frac{N\, \hbar^2}{2\pi}\left(\frac{e}{N\, \hbar}\right)^{2N}\qquad, \quad N\gg 1
\label{eq:gap-width}
\end{eqnarray}
Note that this has the same form as the non-perturbative effects in the large-$N$ expansion of the Gross-Witten-Wadia unitary matrix model, in the strong coupling phase \cite{gw,wadia,goldschmidt}.

These strong-coupling results are very different from the weak-coupling expressions described in Section \ref{sec:weak}. The expansions are convergent rather than divergent, and the expansions do not appear to have a trans-series structure. Nevertheless, we show below in Section  \ref{sec:summary} that in fact a more natural way to interpret the expressions (\ref{eq:conv}) is as trans-series \cite{bdu2}.

\subsection{Theorems Concerning Band and Gap Widths}
\label{sec:theorems}

Here we summarize the known rigorous estimates of the widths of bands and gaps (stability or instability regions) for the Mathieu problem. These generalize to the more general case of a periodic potential with a unique minimum in each period.

\begin{quote}
{\bf Band-Width Theorem} [Harrell \cite{harrell}, Weinstein-Keller \cite{keller}, Connor et al \cite{connor}]: For a periodic potential with a unique minimum in each period, the width of the $N^{\rm th}$ band has the leading behavior
\begin{eqnarray}
\Delta u_{\rm band} &\sim& \frac{2}{\pi}\frac{\partial u}{\partial N}\exp\left[-\frac{1}{\hbar}\, S\right]\left(1+O(\hbar)\right)
\nonumber\\
S&=&\int_{\rm turning \, points} \sqrt{V(x)-V_{\rm min}} \, dx
\label{eq:band-theorem}
\end{eqnarray}
where the exponent involves the action under the barrier. For the Mathieu equation, this produces the leading part of expression (\ref{eq:band-width}), as we discuss in Section \ref{sec:wkb}. Our new result (\ref{eq:one}, \ref{eq:prefactor}) refines this estimate considerably, giving all the $O(\hbar)$ corrections, and moreover expresses these corrections entirely in terms of the perturbative series.
\end{quote}

\begin{quote}
{\bf Gap-Width Theorem} [Dykhne \cite{dykhne}, Weinstein-Keller \cite{keller}, Connor et al \cite{connor}, Avron and Simon \cite{avron}]: For a periodic potential with a unique minimum in each period, the width of the $N^{\rm th}$ gap for $u\gg V_{\rm max}$ has the leading behavior:
\begin{eqnarray}
\Delta u_{\rm gap} &\sim& \frac{2}{\pi}\frac{\partial u}{\partial N}\, \exp\left[-\frac{1}{\hbar} {\rm Im}\, \tilde{S}\right] \left(1+O(\hbar)\right)
\nonumber\\
\tilde{S}&=&\int_{\rm complex\, turning \, points} \sqrt{V(x)-V_{\rm min}} \, dx
\label{eq:gap-theorem}
\end{eqnarray}
where the exponent now involves the action on a cycle connecting complex turning points. For the Mathieu equation, this produces the leading part of expression (\ref{eq:gap-width}), as we discuss in Section \ref{sec:wkb}. 
\end{quote}
Note the remarkable fact  that these leading expressions (\ref{eq:band-theorem}, \ref{eq:gap-theorem}) for the band width and gap width have  a common form, just with different turning points. This will be explained in Section \ref{sec:wkb}. This gap-width result can also be extended, along the lines of  (\ref{eq:one}, \ref{eq:prefactor}) for the bands  \cite{bdu2}.

\section{Uniform WKB  for the Mathieu Equation}
\label{sec:uniform}

In this Section we outline our first approach to the spectrum, which demonstrates how a trans-series expression such as (\ref{trans}) arises for the energy levels, in the weak coupling regime where $\hbar N\ll 1$. Here we outline the overall strategy and summarize a few technical details. For further details, see \cite{Dunne:2014bca,Dunne:2013ada}. The resurgent structure of the Mathieu equation system has also been studied in  \cite{Misumi:2015dua}.

\subsection{Strategy of  Uniform WKB}

Uniform WKB is a well-known approach  \cite{langer,cherry,miller-good}, based on the simple idea of mapping the Schr\"odinger problem to a known "comparison problem". Since the wells are harmonic, the relevant comparison problem is the harmonic oscillator system, whose eigenfunctions are parabolic cylinder functions. 
The novelty of \cite{Dunne:2014bca,Dunne:2013ada} is that we push this expansion well beyond the leading terms studied in  \cite{langer,cherry,miller-good}, which reveals a surprising new layer of rich mathematical structure. See \cite{alvarez,Dunne:2014bca,Dunne:2013ada} for a related approach to the symmetric double-well potential, which also has (degenerate) harmonic minima.

The comparison mapping is achieved by making  a parabolic uniform WKB ansatz for the wave-function: 
\begin{eqnarray}
\psi(x) = \frac{1}{\sqrt{\varphi^\prime(x)}}\, D_\nu \left(\frac{1}{\sqrt{\hbar}}\varphi (x)\right)
\label{uniform}
\end{eqnarray}
Here  $D_\nu$ is the parabolic cylinder function \cite{ww,carl-book}, and $\nu$ is an ansatz parameter whose physical meaning will become clear below. 
It is closely related to the band label $N$.  
This uniform WKB ansatz converts the linear 
Schr\"odinger equation for $\psi(x)$ into  a nonlinear equation for the argument function $\varphi (x)$,  which can then be solved perturbatively. Straightforward local analysis near the (harmonic) potential minimum leads to a perturbative expansion of the energy:
\begin{eqnarray}
u=u_{\rm pert}(\hbar, \nu)=\sum_{n=0}^\infty \hbar^{n} u_n(\nu)
\label{eq:pert}
\end{eqnarray}
The coefficient $u_n(\nu)$ is a polynomial, of degree $n$, in the ansatz parameter $\nu$. In the $\hbar\to 0$ limit, the ansatz parameter $\nu$ tends to an integer $N$,  labelling the unperturbed harmonic oscillator energy level. In fact, when $\nu=N$, the expansion (\ref{eq:pert}) coincides precisely with the result $u_{\rm pert}(\hbar, N)$ of  Rayleigh-Schr\"odinger perturbation theory in (\ref{eq:perturbation}).
As mentioned in the Introduction, this perturbative series expression is incomplete, and indeed ill-defined, because the series is not Borel summable. This is not particularly surprising because so far the analysis has been purely {\it local}, making no reference to the existence of neighboring degenerate minima of the potential. To determine the energy spectrum properly we must impose a {\it global} boundary condition that relates one minimum to another.

This global condition implies that $\nu$ is only {\it exponentially close} to the integer $N$, with a small correction that is a function of both $N$ and $\hbar$ \cite{Dunne:2014bca,Dunne:2013ada}:
\begin{eqnarray}
\nu_{\rm global}(\hbar, N)=N+\delta\nu(\hbar, N)
\label{global}
\end{eqnarray}
The correction term $\delta\nu(, N)$ has a trans-series form:
\begin{eqnarray}
\delta\nu(\hbar, N)=\sum_{k=1}^\infty \sum_{n=0}^\infty  \sum_{l=1}^{k-1} d_{k, n,  l}(N) 
\left(\frac{1}{\hbar^{N+1/2}}\,\exp\left[-\frac{S}{\hbar}\right]\right)^k 
\left(\ln \left[-\frac{1}{\hbar}\right]\right)^l \hbar^{n}
\label{eq:deltanu}
\end{eqnarray}
As explained below, this trans-series structure follows directly from the properties of the parabolic cylinder functions, and it is therefore generic for potentials with degenerate harmonic minima. 

The global boundary condition  determines the ansatz parameter $\nu$ as a function of $\hbar$ and $N$, as in (\ref{global}, \ref{eq:deltanu}), and then the trans-series form of the energy eigenvalue follows immediately by inserting this into the formal perturbative series expansion: 
\begin{eqnarray}
u_{\rm trans}(N, \hbar)=u_{\rm pert}\left(\hbar, N+\delta\nu(N, \hbar)\right)=\sum_{n=0}^\infty \hbar^{n} u_n(N+\delta\nu(N, \hbar))
\label{answer}
\end{eqnarray}
Recall that  the  coefficients $u_n(\nu)$ are polynomials. We are using the fact that a formal expansion of a trans-series is itself a trans-series.

This uniform WKB approach explains in very elementary terms why the  energy generically  has a trans-series form for potentials with degenerate harmonic classical minima: all properties of the $\hbar\to 0$ limit are simply mapped to the resurgent asymptotic properties of the parabolic cylinder functions. All analytic continuations needed to analyze questions of resurgence and cancellation of ambiguities are ultimately expressed in terms of the well-known analytic continuation properties of the parabolic cylinder functions.

\subsection{Perturbative Expansion of the Uniform WKB Ansatz}

The parabolic cylinder function $D_\nu(z)$ satisfies the differential equation
\begin{eqnarray}
\frac{d^2}{dz^2}D_\nu(z)+\left(\nu+\frac{1}{2}-\frac{z^2}{4}\right)D_\nu(z)=0
\label{pcf}
\end{eqnarray}
Therefore the uniform WKB ansatz (\ref{uniform}) converts the linear Schr\"odinger equation (\ref{eq:mathieu}) to a non-linear equation for the argument function $\varphi(x)$:
\begin{eqnarray}
V(x)-u-\frac{1}{8}\varphi^2(\varphi^\prime)^2+\frac{\hbar}{2} \left(\nu+\frac{1}{2}\right)(\varphi^\prime)^2+\frac{\hbar^2}{4}\sqrt{\varphi^\prime}\left(\frac{\varphi^{\prime\prime}}{(\varphi^\prime)^{3/2}}\right)^\prime=0
\label{nle}
\end{eqnarray}
Notice that the ansatz parameter $\nu$ appears in this nonlinear equation, whereas it does not appear in the original Schr\"odinger equation.
This nonlinear equation can be solved using simultaneous perturbative expansions for $\varphi(x)$ and $u$:
\begin{eqnarray}
u&=& u_0+\hbar u_1+\hbar^2 u_2+\dots
\label{pe1}\\
\varphi(x)&=& \varphi_0(x)+\hbar\, \varphi_1(x)+\hbar^2\, \varphi_2(x)+\dots
\label{eq:p2}
\end{eqnarray}
Note that the expansion coefficients $u_n$ and $\varphi_n(x)$  also depend on the ansatz parameter $\nu$, simply because they come from solving the equation (\ref{nle}), in which the parameter $\nu$  now appears.

\subsubsection{Leading Order of the Perturbative Expansion}
The leading  order of (\ref{nle})  determines $\varphi_0(x)$ and $u_0$:
\begin{eqnarray}
\varphi_0^2(\varphi_0^\prime)^2=8(V(x)-u_0)
\label{zeroth}
\end{eqnarray}
It is notationally simpler in this case to shift the coordinate $x$ by $\pi$ so that the minimum of the potential is at $x=0$. Thus we take $V(x)=-\cos x$. Clearly the spectrum is unchanged by this shift. Requiring regular behavior at $x=0$ we determine
\begin{eqnarray}
u_0&=&-1 \\
\varphi_0(x)&=&4\sqrt{2}\, \sin\left(\frac{x}{4}\right)
\label{phi0}
\end{eqnarray}
Physically, we understand  $u_0=-1$ simply as the reference energy of the bottom of the potential well.

\subsubsection{First Order of the Perturbative Expansion}

At $O(\hbar)$, again imposing regularity at $x=0$, we find straightforwardly
\begin{eqnarray}
u_1&=& \nu+\frac{1}{2} \\
\varphi_1(x)&=&\frac{\left(\nu+\frac{1}{2}\right)}{2\sqrt{2}}\frac{\ln \cos \frac{x}{4}}{\sin \frac{x}{4}}
\label{pte}
\end{eqnarray}
Once again, the result $u_1=\nu+\frac{1}{2}$ is clear: it just gives the harmonic energies $\hbar\left(\nu+\frac{1}{2}\right)$, above the reference energy at the bottom of the well: to this order $u(\nu, \hbar)=-1+\hbar\left(\nu+\frac{1}{2}\right)+\dots$.

\subsubsection{Higher Orders of the Perturbative Expansion}

This perturbative expansion can be extended to any desired order. One finds that the energy $u(\nu, \hbar)$ has an expansion of the form
\begin{eqnarray}
u_{\rm pert}(\hbar, \nu)=-1+\hbar \left(\nu+\frac{1}{2}\right)-\sum_{n=2}^\infty \hbar^{n} P_{n}\left(\nu+\frac{1}{2}\right) 
\label{eq:pte}
\end{eqnarray}
where $P_n$ is a polynomial of degree $n$, with definite parity. Explicitly, the first few terms are:
\begin{eqnarray}
u_{\rm pert}(\hbar, \nu)
&\sim& -1 +\hbar \left[\nu+\frac{1}{2}\right] -\frac{\hbar^2}{16}\left[\left(\nu+\frac{1}{2}\right)^2+\frac{1}{4}\right]-\frac{\hbar^3}{16^2}\left[\left(\nu+\frac{1}{2}\right)^3+\frac{3}{4}\left(\nu+\frac{1}{2}\right)\right] \nonumber\\
&&  - \frac{\hbar^4}{16^3}\left[ \frac{5}{2}\left(\nu+\frac{1}{2}\right)^4+\frac{17}{4}\left(\nu+\frac{1}{2}\right)^2+\frac{9}{32}\right]  \nonumber\\
&& 
-  \frac{\hbar^5}{16^4}\left[ \frac{33}{4}\left(\nu+\frac{1}{2}\right)^5+\frac{205}{8}\left(\nu+\frac{1}{2}\right)^3+\frac{405}{64}\left(\nu+\frac{1}{2}\right)\right] -\dots
\label{eq:weak2}
\end{eqnarray}
Replacing $\nu$ by an integer quantum number $N$, this expansion (\ref{eq:weak2}) agrees with the result (\ref{eq:weak}) of Rayleigh-Schr\"odinger perturbation theory about the $N^{\rm th}$ harmonic oscillator level. But, as before, this is a formal, divergent asymptotic expansion, which moreover is not Borel summable, so this is not the whole story. We are missing the information about the fact that the spectrum consists of {\it bands}, not of discrete levels. This omission is because we have not yet applied the correct physical boundary conditions.

\subsection{Global Boundary Condition: Connecting Different Minima}
\label{sec:connecting}

To fully determine the spectrum, we need to extend the previous {\it local} analysis, in the neighborhood of a minimum of the potential, with {\it global} information about how neighboring minima are related.

It is well known (see Figure \ref{fig:fig1}) that each perturbative level labeled by the integer index $N$ splits into a continuous band of states (in classical language these are the ``stability regions'' of the Mathieu equation). In the QM Schr\"odinger context this phenomenon arises from the Bloch condition: 
\begin{eqnarray}
\psi(x+2\pi)=e^{i\,\theta}\psi(x)
\label{eq:bloch}
\end{eqnarray}
where $\theta$ is a real angular parameter that labels states within a given band of the spectrum. Conventional Floquet analysis \cite{magnus,kenbook,nist-mathieu,ww,meixner} expresses this Bloch boundary condition in terms of the discriminant. Define two independent solutions $\psi_{1}(x)$ and $\psi_{2}(x)$, normalized as follows at some arbitrary chosen point (which we take here to be at $x=-\pi$, the center of a barrier between two potential minima: recall that here we are taking the potential $V(x)=-\cos x$):
\begin{eqnarray}
\begin{pmatrix}
\psi_1\left(-\pi\right) & \psi_1^\prime \left(-\pi\right) \cr
\psi_2\left(-\pi\right) & \psi_2^\prime \left(-\pi\right)
\end{pmatrix}
=
\begin{pmatrix}
1& 0\cr
0 & 1
\end{pmatrix}
\label{basis}
\end{eqnarray}
Then, using the symmetry of the Mathieu potential, we can re-write the Bloch condition in the compact form \cite{nist-mathieu}:
\begin{eqnarray}
\cos\theta
&=&
\psi_1(\pi)
\label{bloch}
\end{eqnarray}
This is the {\it exact quantization condition}, which determines the band spectrum. To make this exact quantization condition practically useful,  we need a way to evaluate the right-hand-side of (\ref{bloch}). This can be done numerically, or using asymptotics, as we do in this paper.
The band-edge states correspond to $\theta=0$ or $\theta=\pi$, and are periodic or anti-periodic functions, respectively. For other values of $\theta$ the wavefunctions are quasi-periodic (see (\ref{eq:bloch})), corresponding to states within a given band. All this is very well known. 

Now let us apply this global boundary condition (\ref{bloch}) to the uniform WKB ansatz (\ref{uniform}).
First, define even and odd functions 
\begin{eqnarray}
f_{\pm}(x)&=&\frac{1}{\sqrt{\varphi^\prime(x)}}\left(D_\nu\left(\frac{\varphi(x)}{\sqrt{\hbar}}\right) \pm D_\nu\left(-\frac{\varphi(x)}{\sqrt{\hbar}}\right)\right) 
\label{eq:pm}
\end{eqnarray}
where we  have used the fact that  $\varphi(x)$ is odd, and $\varphi^\prime(x)$ is even, on the interval $x\in \left[-\pi, \pi \right]$. 
Then the normalized basis functions in (\ref{basis}) are
\begin{eqnarray}
\psi_1(x)&=& \frac{1}{{\mathcal W}}\left(f_-^\prime\left(-\pi\right)\, f_+\left(x\right) - f_+^\prime\left(-\pi\right)\, f_-(x)\right) 
\label{psi1}\\
\psi_2(x)&=& \frac{1}{{\mathcal W}}\left(-f_-\left(-\pi\right)\, f_+(x) + f_+\left(-\pi\right)\, f_-(x)\right) 
\label{psi2}
\end{eqnarray}
where ${\mathcal W}$ is the (constant) Wronskian:
\begin{eqnarray}
{\mathcal W}\equiv f_+(x)\, f_-^\prime(x)-f_+^\prime(x)\, f_-(x)=-\sqrt{\frac{8\pi}{\hbar}}\frac{1}{\Gamma(-\nu)}
\label{wronskian}
\end{eqnarray}
Then the exact (Bloch) quantization condition (\ref{bloch}) can be written (in two equivalent forms):
\begin{eqnarray}
\cos\theta
&=&1+\frac{2}{{\mathcal W}}\, f_+^\prime\left(\pi \right)\, f_-\left(\pi \right)
\label{b2}\\
&=&-1+\frac{2}{{\mathcal W}}\, f_-^\prime\left( \pi \right)\, f_+\left( \pi \right) 
\label{b3}
\end{eqnarray}
Thus, the global boundary condition is imposed at the midpoint between two neighboring perturbative vacua: $x_{\rm midpoint}=\pi$. Moreover, the global condition is expressed in terms of  parabolic cylinder functions evaluated at $x_{\rm midpoint}$. This Bloch condition results in the perturbative energy level splitting into a continuous band, with states within the band labeled by the angular parameter $\theta$. The bottom of the lowest band has $\theta=0$ and its wave function is an even function, while the top of the lowest band has $\theta=\pi$ and its wave function is an odd function. For such band-edge states the Bloch conditions (\ref{b2}, \ref{b3}) take a simpler form:
\begin{eqnarray}
({\rm even \, state}) &:& f_-\left(\pi\right)=0
\label{sg-lower}\\
({\rm odd \, state}) &:& f_+\left(\pi\right)=0
\label{sg-upper}
\end{eqnarray}
Using the uniform WKB ansatz (\ref{eq:pm}) we find that these band edges are determined by the exact condition:
\begin{eqnarray}
D_\nu\left(\frac{\varphi(\pi)}{\sqrt{\hbar}}\right)= \pm D_\nu\left(-\frac{\varphi(\pi)}{\sqrt{\hbar}}\right)
\label{eq:bloch3}
\end{eqnarray}

Now we come to the main point. The argument $\varphi(\pi)$, just like the perturbative expansion of the energy in (\ref{eq:weak2}),  is a Borel non-summable series in $\hbar$, with leading term being non-zero: $\varphi_0(\pi)=4$. Thus, as $\hbar\to 0$ we are interested in the asymptotics of the parabolic cylinder functions at large magnitude of their argument. But in order to make sense of the Borel non-summable series we must analytically continue in $\hbar$ off the Stokes line, the positive real axis, $\hbar>0$. Then the argument of the parabolic cylinder functions in (\ref{eq:bloch3}) becomes complex, and we need to use the asymptotics of these functions off the real axis \cite{nist-mathieu,ww}:
\begin{eqnarray}
D_\nu(z)\sim z^\nu\, e^{-z^2/4}\left(1+\dots \right)+e^{\pm i\pi\nu} \frac{\sqrt{2\pi}}{\Gamma(-\nu)} z^{-1-\nu}\, e^{z^2/4} \left(1+\dots \right)
\qquad , \qquad \frac{\pi}{2} < \pm \,{\rm arg}(z) <\pi
\label{full}
\end{eqnarray}
The sign here depends on which way we continue off the real axis. The ellipsis denote known fluctuation terms \cite{nist-mathieu,ww,Dunne:2014bca}. Thus, we see that the exact quantization condition for the band edges (\ref{eq:bloch3}) requires a balance between the two exponential terms in (\ref{full}). Had we been on the real axis, we would eliminate the growing exponential in (\ref{full})  by setting $\nu$ equal to a non-negative integer, but this is just perturbation theory.
Balancing the two exponentials expresses the exact quantization condition in the following form:
\begin{eqnarray}
\frac{\left(e^{\pm i\pi}\, \hbar\right)^\nu}{\Gamma(-\nu)}=\pm \left(\varphi(\pi)\right)^{2\nu+1}\frac{e^{-\frac{1}{2} \frac{\varphi^2(\pi)}{\hbar}}}{\sqrt{2\pi\, \hbar}}\left(1+\dots\right)
\label{eq:bloch4}
\end{eqnarray}
Expanding about an integer, writing $\nu=N+\delta \nu$, and the fact that $\frac{\varphi^2(\pi)}{2\hbar}= \frac{8}{\hbar} -\left(\nu+\frac{1}{2}\right) \ln 2+\dots$, we find the exponentially small shift in (\ref{global}).
The higher order terms generate  powers of $\ln \hbar$ (for further details, see \cite{Dunne:2014bca}). Thus we see immediately how the trans-series structure arises: inserting $\nu=N+\delta\nu (\hbar, N)$ into the formal expansion (\ref{eq:pert}, \ref{eq:weak2}), we generate a trans-series expression for $u$, of the form (\ref{trans}). By construction, all analytic continuations are self-consistent, because we have simply used the analytic continuation properties of the parabolic cylinder functions.  All higher fluctuation terms can be deduced straightforwardly  from the fluctuations of the parabolic cylinder functions.

This {\it resurgence} procedure works because the global boundary condition is expressed in terms of parabolic cylinder functions of $1/\hbar$,  and for these functions the analytic continuation properties are rigorously known. Thus, trans-series expressions of the form (\ref{trans}) also arise for other potentials with harmonic minima. Ambiguities only arise if you first expand the global boundary condition and look at just one isolated portion of the resulting trans-series expansion, for example just the perturbative part, or just some particular multi-instanton sector. When viewed as a whole, the analytic continuation of the trans-series expression is unique and exact, and leads to real and unambiguous energy eigenvalues.

\section{Relating  Perturbative and Non-Perturbative Sectors}
\label{sec:magic}

\subsection{Results of Zinn-Justin and Jentschura}
\label{sec:zjj}

We  now compare with the beautiful  results of Zinn-Justin and Jentschura (ZJJ) \cite{ZinnJustin:2004ib}. (Recall Eqs (\ref{zj}, \ref{hbar-g}), which translate between their notation for the Mathieu equation and ours.) ZJJ conjectured an exact quantization condition of the form:
\begin{eqnarray}
\left(\frac{32}{\hbar}\right)^{-B_{\rm ZJJ}}\frac{e^{\frac{1}{2}A_{\rm ZJJ}}}{\Gamma\left(\frac{1}{2}-B_{\rm ZJJ}\right)}+
\left(-\frac{32}{\hbar}\right)^{-B_{\rm ZJJ}}\frac{e^{-\frac{1}{2}A_{\rm ZJJ}}}{\Gamma\left(\frac{1}{2}+B_{\rm ZJJ}\right)}=\frac{2\, \cos\theta}{\sqrt{2\,\pi}}
\label{eq:zjj-exact}
\end{eqnarray}
where $\theta$ is the Bloch angle, and
$A_{\rm ZJJ}(\hbar, E)$ and $B_{\rm ZJJ}(\hbar, E)$ are functions of energy that need to be computed. ZJJ derived the following expansions (we rewrite the following expressions in terms of $\hbar \equiv 16 g$ instead of $g$):
\begin{eqnarray}
B_{\rm ZJJ}(\hbar, E)&=&E+\frac{\hbar}{16}\left(\frac{1}{4}+E^2\right) 
+\left(\frac{\hbar}{16}\right)^2 \left(\frac{5E}{4}+3 E^3\right) 
+\left(\frac{\hbar}{16}\right)^3\left(\frac{17}{32}+\frac{35 E^2}{4}+\frac{25 E^4}{2}\right) \nonumber\\
&& +\left(\frac{\hbar}{16}\right)^4\left(\frac{721 E}{64}+\frac{525 E^3}{8}+\frac{245 E^5}{4}\right) +\dots
\label{eq:zjjb}\\
A_{\rm ZJJ}(\hbar, E)&=&\frac{16}{\hbar}+\frac{\hbar}{16} \left(\frac{3}{4}+3 E^2\right)
+\left(\frac{\hbar}{16}\right)^2 \left(\frac{23 E}{4}+11 E^3\right) 
 +\left(\frac{\hbar}{16}\right)^3 \left(\frac{215}{64}+\frac{341 E^2}{8}+\frac{199 E^4}{4}\right)\nonumber\\
 &&+\left(\frac{\hbar}{16}\right)^4 \left(\frac{4487 E}{64}+326 E^3+\frac{1021 E^5}{4}\right)+\dots
 \label{eq:zjja}
 \end{eqnarray}
ZJJ showed that expanding the transcendental quantization condition (\ref{eq:zjj-exact}) leads to a trans-series expression for the energy eigenvalue, and made extensive numerical checks of the associated resurgent cancellations. We also note that the computation of $B_{\rm ZJJ}(\hbar, E)$ is straightforward, while $A_{\rm ZJJ}(\hbar, E)$ is considerably more involved  \cite{ZinnJustin:2004ib}.
 
\subsection{Inversion and a Surprise}
 \label{sec:inversion}
 
Consider inverting the expression for $B_{\rm ZJJ}(E, \hbar)$, to express it as $E_{\rm  ZJJ}(\hbar, B)$  \cite{Dunne:2013ada,Dunne:2014bca}:
\begin{eqnarray}
E_{\rm  ZJJ}(\hbar, B)&=&
B-\frac{\hbar}{16}\left(B^2+\frac{1}{4}\right)
-\left(\frac{\hbar}{16}\right)^2\left(B^3+\frac{3 B}{4}\right)
-\left(\frac{\hbar}{16}\right)^3\left(\frac{5B^4}{2}+\frac{17 B^2}{4}+\frac{9}{32}\right)\nonumber\\
   && -\left(\frac{\hbar}{16}\right)^4\left(\frac{33 B^5}{4}+\frac{205 B^3}{8}+\frac{405 B}{64}\right) -\dots
\label{eq:bsg}
\end{eqnarray}
We see that this agrees with the perturbative weak coupling  expansion (\ref{eq:weak}), with the definitions (\ref{hbar-g})  and  the identification of $B$ with the band label number $N$:
\begin{eqnarray}
B= N+\frac{1}{2} 
\label{eq:b}
\end{eqnarray}
Thus, computing  the function $B_{\rm ZJJ}(\hbar, E)$ is equivalent to computing 
Rayleigh-Schr\"odinger perturbation theory. These are the fluctuations about the perturbative vacuum.

Using (\ref{eq:bsg}), the non-perturbative function $A_{\rm ZJJ}(\hbar, E)$  in (\ref{eq:zjja}) can be re-expressed as a function of $B$:
\begin{eqnarray}
 A_{\rm ZJJ}(\hbar, B)&=&\frac{16}{\hbar}+\frac{\hbar}{16}\left(3 B^2+\frac{3}{4}\right) 
 +\left(\frac{\hbar}{16}\right)^2\left(5 B^3+\frac{17 B}{4}\right)
 + \left(\frac{\hbar}{16}\right)^3 \left(\frac{55 B^4}{4}+\frac{205 B^2}{8}+\frac{135}{64}\right) \nonumber\\
&& +\left(\frac{\hbar}{16}\right)^4\frac{9}{64} \left(336 B^5+1120 B^3+327 B\right)    +\dots
\label{eq:asg}
\end{eqnarray}
Notice the striking similarities between terms in the expansions of $A_{\rm ZJJ}(\hbar, B)$ and $E_{\rm ZJJ}(\hbar, B)$. To see this explicitly, we compute:
\begin{eqnarray}
\frac{\partial E_{\rm ZJJ}( \hbar, B)}{\partial B}&=&
1-\frac{\hbar}{8} B
-\left(\frac{\hbar}{16}\right)^2\left(3 B^2+\frac{3}{4}\right)
-\left(\frac{\hbar}{16}\right)^3\left(10 B^3+\frac{17 B}{2}\right)
\nonumber\\&& 
   -\left(\frac{\hbar}{16}\right)^4\left(\frac{165 B^4}{4}+\frac{615 B^2}{8}+\frac{405}{64}\right) -\dots
\label{eq:dedb}
\end{eqnarray}
and compare this with
   \begin{eqnarray}
 -\frac{\hbar^2}{16} \frac{\partial A_{\rm ZJJ}(\hbar, B)}{\partial \hbar}&=& 1
 -\left(\frac{\hbar}{16}\right)^2\left(3 B^2+\frac{3}{4}\right)
-\left(\frac{\hbar}{16}\right)^3\left(10 B^3+\frac{17 B}{2}\right)
\nonumber\\   &&
    -\left(\frac{\hbar}{16}\right)^4\left(\frac{165 B^4}{4}+\frac{615 B^2}{8}+\frac{405}{64}\right) -\dots 
  \label{eq:dadg}
     \end{eqnarray}
We deduce the remarkably simple relation:
\begin{eqnarray}
\frac{\partial E_{\rm ZJJ}}{\partial B}=- \frac{\hbar}{16}\left(2B+\hbar\frac{\partial A_{\rm ZJJ}}{\partial \hbar}\right)
\label{eq:magic}
\end{eqnarray}
It is straightforward to check this to higher orders in $\hbar$.

In other words, the non-perturbative  function $A_{\rm ZJJ}(\hbar, B)$, and therefore also $A_{\rm ZJJ}(\hbar, E)$,  can be deduced immediately from knowledge of the perturbative energy $E_{\rm ZJJ}(\hbar, B)$. Thus, only one of the two functions $B_{\rm ZJJ}(\hbar, E)$ and $A_{\rm ZJJ}(\hbar, E)$ is actually needed to generate the entire trans-series.  Note that this fact is not at all obvious when these functions are written in terms of $E$ as $B_{\rm ZJJ}(\hbar, E)$ and $A_{\rm ZJJ}(\hbar, E)$, as in (\ref{eq:zjjb}, \ref{eq:zjja}), but become clear when everything is expressed in terms of $B$, as in (\ref{eq:bsg}, \ref{eq:asg}). We will understand the reason for this in Section \ref{sec:wkb}. The implication of this fact is that the exact quantization condition (\ref{eq:zjj-exact}) involves only one function. See for example Eq. (68) in \cite{Dunne:2014bca}.

\subsection{Implications and Two Independent Confirmations}
\label{sec:confirm}

This relation between the $A_{\rm ZJJ}$ and $B_{\rm ZJJ}$ functions has some remarkable implications. Recall that
the function $E_{\rm ZJJ}(\hbar, B)$ describes the fluctuations about the perturbative vacuum, while  $A_{\rm ZJJ}(\hbar, B)$ describes the fluctuations about the single-instanton \cite{ZinnJustin:2004ib,Dunne:2013ada}. For example,
the single-instanton fluctuation factor comes from
\begin{eqnarray}
\frac{\partial E_{\rm ZJJ}}{\partial B}\, e^{-\frac{1}{2} A_{\rm ZJJ}}&\sim& e^{-8/\hbar} \left(1-  \frac{\hbar}{8} B - ...\right)\left(1-\frac{\hbar}{32}\left(3B^2+\frac{3}{4}\right)- ...\right) \nonumber\\
&=&e^{-8/\hbar} \left(1-\frac{\hbar}{32}\left(3B^2+4B+\frac{3}{4}\right) -\dots\right)
\label{eq:sg-fluc}
\end{eqnarray}
in agreement with the fluctuation factor in (\ref{eq:band-width}). At first sight, it looks from (\ref{eq:sg-fluc}) as though one needs {\it both} $E_{\rm ZJJ}(\hbar, B)$ and  $A_{\rm ZJJ}(\hbar, B)$ in order to compute the  fluctuations about the single-instanton sector.
But because of the simple relation (\ref{eq:magic}) between $E_{\rm ZJJ}(\hbar, B)$ and  $A_{\rm ZJJ}(\hbar, B)$, it is enough to know just $E_{\rm ZJJ}(\hbar, B)$, which means just knowing perturbation theory. 

Combining these facts, and re-writing in terms of the energy eigenvalue $u$, we arrive at the result written in the compact form (\ref{eq:one}, \ref{eq:prefactor}). For example, going to the next higher order in $\hbar$ we find:
  \begin{eqnarray}
\Delta u_{\rm band}( \hbar, N)
& \sim&  \frac{2 \hbar}{\sqrt{2 \pi}} \frac{1}{N!} \left(\frac{32}{\hbar}\right)^{N+1/2}
\exp\left[-\frac{8}{\hbar}\right]\left\{ 1-\frac{\hbar}{32}\left[3 \left(N+\frac{1}{2}\right)^2+4 \left(N+\frac{1}{2}\right) +\frac{3}{4}\right] \right.
\label{eq:band-width2} \\
&& \left.+\frac{\hbar^2}{32768}\left(144 \left(N+\frac{1}{2}\right)^4+64 \left(N+\frac{1}{2}\right)^3-312 \left(N+\frac{1}{2}\right)^2-176\left(N+\frac{1}{2}\right)-87\right)+O(\hbar^3)\right\}  \nonumber
\end{eqnarray}
We stress that this is a completely constructive relationship. Given some number of orders of the perturbative expansion $u_{\rm pert}(\hbar, N)$ about the perturbative vacuum, from (\ref{eq:one}, \ref{eq:prefactor}) we can immediately write down the same number of orders of the fluctuations  about the single-instanton sector. In (\ref{eq:band-width2}) we have shown the first few orders of this fluctuation.

Moreover, this relation between perturbative and non-perturbative effects extends throughout the entire trans-series, to all higher multi-instanton sectors \cite{Dunne:2013ada}. They are all encoded in the perturbative series $u_{\rm pert}(\hbar, N)$. Therefore, the fluctuations about the single-instanton saddle, and all other non-perturbative saddles, are precisely encoded in the fluctuations about the perturbative vacuum.   This surprising result  is in fact consistent with the ambitious goal of resurgence, which claims that  the expansion about one saddle contains, in principle, information about the expansions around other saddles, provided one knows how different saddles are connected. This connection is provided by the exact quantization condition (\ref{bloch}), which is itself a statement of the Bloch boundary condition \cite{Dunne:2013ada}.

At the one-instanton level, we have two independent confirmations of this result. Long ago, in a truly remarkable paper, Dingle and M\"uller computed many orders of the  series corrections to the exponential band splitting, directly from the asymptotics of the Mathieu differential equation \cite{muller}. We can compare (\ref{eq:band-width2}) with equation (72) of \cite{muller}, and after adjusting notation (their $q_0\equiv (2N+1)$, and their $h\equiv 2/\hbar$) we see complete  agreement. This comparison can easily be carried out to higher orders.

From the physics perspective, an even more interesting comparison is to a recent direct brute-force Feynman diagram computation of the fluctuations about the single-instanton sector, to three loop order \cite{Escobar-Ruiz:2015rfa} (the corresponding result has also been confirmed for the double-well system \cite{Escobar-Ruiz:2015rfa}). Specifically, for the single-instanton expansion of the lowest ($N=0$) band, the diagrammatic computation found the fluctuation factor (translated to our normalizations: their $g\equiv \hbar/8$):
\begin{eqnarray}
\exp\left[-\frac{8}{\hbar}\right] \left[1{\color{blue}-\frac{7}{8}} \left(\frac{\hbar}{8}\right) {\color{blue}-0.460937498}\left(\frac{\hbar}{8}\right)^2  -\dots\right]
\label{eq:shuryak}
\end{eqnarray}
This diagrammatic computation involved more than 20 three-loop Feynman digrams, each involving propagators in the presence of an instanton, and only some of these diagrams could be evaluated exactly. Hence the third coefficient is only known numerically.
This field-theoretic result should be compared with  the fluctuation factor in (\ref{eq:band-width2}) with $N=0$:
\begin{eqnarray}
\exp\left[-\frac{8}{\hbar}\right] \left[1{\color{blue}-\frac{7}{8}} \left(\frac{\hbar}{8}\right){\color{blue}- \frac{59}{128}}\left(\frac{\hbar}{8}\right)^2 -\dots\right]
\label{eq:shuryak2}
\end{eqnarray}
We see agreement up to 7 decimal places. Considering the complexity of the Feynman diagrammatic approach \cite{Escobar-Ruiz:2015rfa}, compared with the simplicity of the result (\ref{eq:prefactor}), we see that resurgence is telling us something quite novel and non-trivial about perturbation theory.  Furthermore, we note that in the diagrammatic computation, certain diagrams evaluated to numbers involving zeta values, but these all presumably cancel in the end because the final coefficient is rational. This is surprisingly reminiscent of such cancellations found in QFT at higher loops, a fact that has fascinating number theoretic and combinatorial implications \cite{Broadhurst:1995dq}. 

It is also interesting to compare the result (\ref{eq:one}, \ref{eq:prefactor})  with the theorems of Harrell, Weinstein-Keller and Connor et al \cite{keller,harrell,connor} listed in Section \ref{sec:theorems}. These results give one part of the prefactor to leading order, namely the density of states factor $\partial u/\partial N$, but do not say anything about higher orders in $\hbar$. Presumably there is a strong theorem here awaiting a more rigorous proof.

\section{All-orders WKB Analysis of the Mathieu Equation: Actions and Dual Actions}
\label{sec:wkb}

There is another, complementary,  WKB approach to this problem, often called ``exact WKB'', or ``all-orders WKB'' \cite{howls-book}. While the uniform WKB approach has the advantage that it explains straightforwardly where the trans-series structure comes from in the weak-coupling spectral region, the all-orders WKB has an advantage that it can easily describe both high and low energy, and even energies near the top of the barrier. There is also an interesting connection to low energy beahvior of certain supersymmetric quantum field theories \cite{Nekrasov:2009rc,Mironov:2009uv,He:2010xa,Huang:2011qx,KashaniPoor:2012wb,Krefl:2013bsa,Gorsky:2014lia,Basar:2015xna}.

We begin with Dunham's  all-orders WKB action \cite{dunham,carl-book}:
\begin{eqnarray}
a(\hbar, u)&=& \frac{\sqrt{2}}{2\pi}\left(\oint_C \sqrt{u-V} dx-\frac{\hbar^2}{2^6}\oint_C \frac{(V^\prime)^2}{(u-V)^{5/2}} dx 
-\frac{\hbar^4}{2^{13}}\oint_C \left(\frac{49 (V^\prime)^4}{(u-V)^{11/2}}- \frac{16 V^\prime V^{\prime\prime\prime}}{(u-V)^{7/2}}\right)dx-\dots \right)
\label{eq:dunham}
\end{eqnarray}
where the contour integral encircles the turning points. The leading term is just the familiar WKB action.
We write this expansion as a formal series in powers of $\hbar^2$, with coefficients that are functions of the energy $u$:
\begin{eqnarray}
 a(\hbar, u)=\sum_{n=0}^\infty \hbar^{2n} \, a_n(u)
 \label{eq:allorders}
 \end{eqnarray}
Remarkably, the higher-order WKB actions can be obtained by acting on these leading WKB actions with  differential operators with respect to the energy $u$ \cite{Mironov:2009uv,He:2010xa,Basar:2015xna}. This follows from the fact that for $V=\cos(x)$, the numerators in \eqref{eq:dunham}, which are given by the derivatives of $V$, can be re-expressed as polynomials of $V$. Therefore by differentiating $\sqrt{u-V}$ with respect to $u$, and taking appropriate combinations, one can generate the integrands in \eqref{eq:dunham} up to total derivatives which vanish after integrating around the turning points. For example, at the next two orders:
\begin{eqnarray}
a_1(u)&=&{1\over 48}\left(2u{d^2\over du^2}+{d\over du}\right) a_0(u)\\
a_2(u)&=&{1\over 2^9 45}\left(28 u^2{d^4\over du^4}+120u{d^3\over du^3}+75{d^2\over du^2}\right) a_0(u)
\label{eq:ans}
\end{eqnarray}
In fact, $a_0(u)$ satisfies a 2nd-order Picard-Fuchs equation (\ref{eq:pf0}), so these can be further simplified, but we do not need this fact here.

The all-orders WKB expression for the {\it location} of  a narrow band or gap can be expressed as \cite{dunham,carl-book}:
\begin{eqnarray}
a(\hbar, u) =
\begin{cases}
 \frac{1}{2} \, N\, \hbar \qquad\qquad \quad ({\rm gap}) \cr
 \frac{1}{2}  \left(N+\frac{1}{2}\right) \hbar \qquad ({\rm band})
\end{cases}
\label{eq:bs}
\end{eqnarray}
This ``all-orders Bohr-Sommerfeld'' formula relates a complicated expression in the energy, $u$, and coupling, $\hbar$, to an integer that labels the band or gap. This can be inverted to express the energy as a function of $N$ and $\hbar$.  For example, when $u\approx -1$, we can expand and invert, which leads to the perturbative expansion (\ref{eq:weak}), as we illustrate in the next sub-section. Alternatively, we can expand at $u\gg 1$ and then invert, which leads to the strong-coupling  expansion (\ref{eq:strong}), as we illustrate in the following sub-section. Expanding  near the barrier top, $u\approx 1$, yields information about the spectrum in this regime.

All this is perturbative information, obtained from the formal all-orders WKB expansion (\ref{eq:dunham}) in powers of $\hbar$.
But, as is clear from the Mathieu spectrum shown in Figure 1, we are also interested in non-perturbative information related to the width of the bands and gaps, whose central locations are given by (\ref{eq:bs}). This non-perturbative information can be extracted  from the same all-orders expansion (\ref{eq:dunham}), but with different contours, now encircling the other turning points; for example, for the energy range below the barrier top, the new turning points are those associated with tunneling through the barrier between two adjacent wells. For energies above the barrier top, these turning points merge and go into the complex plane, and this expression continues smoothly. Thus, we write this ``dual'' action as
\begin{eqnarray}
a^D(\hbar, u)&=& \frac{\sqrt{2}}{2\pi}\left(\oint_{C_D} \sqrt{u-V} dx-
\frac{\hbar^2}{2^6}\oint_{C_D} \frac{(V^\prime)^2}{(u-V)^{5/2}} dx 
-\frac{\hbar^4}{2^{13}}\oint_{C_D} \left(\frac{49 (V^\prime)^4}{(u-V)^{11/2}}- \frac{16 V^\prime V^{\prime\prime\prime}}{(u-V)^{7/2}}\right)dx-\dots \right)
\label{eq:dunham2}
\end{eqnarray}
integrated over the dual integration cycle $C_D$, which encircles these dual turning points. For the Mathieu problem there are just two independent cycles, $C$ and $C_D$,  and they correspond to the generators of the two cycles of the underlying torus. Note  that the higher-order WKB dual actions can also be obtained from the leading one by acting with certain differential operators: in fact, since the only difference is the different cycles, the relations have exactly the same form as in (\ref{eq:ans}), with $a_n$ replaced with $a_n^D$.
With proper analytic continuation, this quantization condition permits smooth transitions and dualities between the various spectral regions, connecting weak and strong coupling, and also the bottom and top of the wells. The distinction between the various regions is encoded in the location of the turning points in the complex plane, and the associated Stokes lines.  For energies inside the wells there are real turning points. As the energy approaches the barrier top,  the turning points  come together and coalesce, and move apart again along the imaginary axis for energy above the barrier top.

Given this dual action, there is an elegant compact formula (\ref{eq:band-theorem}) for the leading width of a {\it band}, for energies well below the potential barrier (i.e. here $-1\leq u\ll 1$). In terms of the action and dual action, this leading order result reads:
\begin{eqnarray}
\Delta u_{\rm band} \sim \frac{2}{\pi}\frac{\partial u}{\partial N}\, e^{-\frac{2 \pi}{\hbar} {\rm Im}\, a_0^D}
\sim \frac{\hbar}{\pi} \frac{\partial u}{\partial a_0}\, e^{-\frac{2 \pi}{\hbar} {\rm Im}\, a_0^D} 
\label{eq:band}
\end{eqnarray}
There is also  an elegant compact formula (\ref{eq:gap-theorem}) for the leading width of a {\it gap}, for energies well above the potential barrier (i.e. here $u\gg1$). In terms of the action and dual action, this leading order result reads:
\begin{eqnarray}
\Delta u_{\rm gap} \sim \frac{2}{\pi}\frac{\partial u}{\partial N}\, e^{-\frac{2 \pi}{\hbar} {\rm Im}\, a_0^D}
\sim \frac{\hbar}{\pi} \frac{\partial u}{\partial a_0}\, e^{-\frac{2 \pi}{\hbar} {\rm Im}\, a_0^D} 
\label{eq:gap}
\end{eqnarray}
Notice that these formula are the same! Furthermore, notice that the prefactor can be interpreted as a frequency, and the exponential factor as a probability of tunneling through the barrier: for band widths, the tunneling is between real turning points, but for gap widths it is tunneling between complex turning points.
The energy dependence of  $a_0$ and $a_0^D$ is very different in the different spectral regions, and this explains the difference between the final expressions (\ref{eq:band-width}, \ref{eq:gap-width})  for the Mathieu band and gap widths, even though they come from expressions with a common  form.

\subsection{Weak Coupling}

The leading order terms of the actions are expressed in terms of elliptic integrals
\begin{eqnarray}
a_0(u)&=&\frac{\sqrt{2}}{\pi}\, \int_{-1}^u dy\, \sqrt{\frac{y-u}{y^2-1}}  
= {4\over \pi} \left({\mathbb E}\left(\frac{1+u}{2}\right)-\frac{1}{2} (1-u)\, {\mathbb K}\left(\frac{1+u}{2}\right)\right)\\
a_0^D(u)&=&- \frac{\sqrt{2}}{\pi}\, \int_{u}^1 dy\, \sqrt{\frac{y-u}{y^2-1}}  
= -{4i\over \pi} \left({\mathbb E}\left(\frac{1-u}{2}\right)-\frac{1}{2} (1+u)\, {\mathbb K}\left(\frac{1-u}{2}\right)\right)\,.
\label{aperiods}
\end{eqnarray}
These actions are two independent solutions of the second-order Picard-Fuchs equation:
\begin{eqnarray}
\frac{d^2 a_0}{du^2}=\frac{1}{4}(1-u^2) a_0(u)
\label{eq:pf0}
\end{eqnarray}
and consequently they satisfy the Wronskian relation
\begin{eqnarray}
a_0(u) \frac{d a_0^D(u)}{du}- a_0^D(u) \frac{d a_0(u)}{du} ={2 i\over \pi} 
\label{eq:picard}
\end{eqnarray}
which simply expresses the Legendre relation [here ${\mathbb K}^\prime$ denotes ${\mathbb K}(1-k^2)$, using the conventions of \cite{nist-mathieu}]:
\begin{eqnarray}
{\mathbb E}{\mathbb K}' +{\mathbb E}' {\mathbb K} -{\mathbb K} {\mathbb K}'=\frac{\pi}{2}\,.
\label{eq:legendre}
\end{eqnarray}

The higher-order WKB actions can be obtained by the action of certain differential operators on $a_0(u)$ or $a_0^D(u)$, as noted above in (\ref{eq:ans}).
For example, the first two next-to-leading order actions calculated from \eqref{eq:ans} are 
     \begin{eqnarray}
a_1(u)&=& \frac{1}{48 \pi  \left(1-u^2\right)}\left((1-u) {\mathbb K}\left(\frac{1+u}{2}\right)+2 u \, {\mathbb E}\left(\frac{1+u}{2}\right)\right)\\
a_2(u)&=&-\frac{1}{46080 \pi \left(1-u^2\right)^3}\left[ (1-u) (4u^3+93 u^2-60 u+75) {\mathbb K}\left(\frac{1+u}{2}\right) \right. \nonumber  \\ 
                 &&\left.+2 \left(4 u^4-153 u^2-75\right) {\mathbb E}\left(\frac{1+u}{2}\right)\right]
   \end{eqnarray}
   and similarly for $a_1^D(u)$ and $a_2^D(u)$.
   
\subsubsection{Band Center Location from Bohr-Sommerfeld at Weak Coupling}

The all-orders Bohr-Sommerfeld expression (\ref{eq:bs}) that identifies $u$ with the center of the band is:
\begin{eqnarray}
\frac{\hbar}{2}\,  \left(N+\frac{1}{2}\right)= a(u, \hbar) \equiv \sum_{n=0}^\infty \hbar^{2n} a_n(u)
\label{eq:bs2}
\end{eqnarray}
The WKB actions $a_n(u)$ can be expanded near the bottom of the wells, $u\sim -1$:
\begin{eqnarray}
a_0(u)&\sim& \frac{u+1}{2}+\frac{ (u+1)^2}{32} +\frac{3  (u+1)^3 }{512}
+\frac{25 (u+1)^4}{16384}+
 \frac{245 (u+1)^5}{524288}+\dots
 \label{eq:dyonic-actions1}
 \\
a_1(u)&\sim&\frac{1}{128}+\frac{5(u+1)}{2048}+ \frac{35 (u+1)^2}{32768} +
\frac{525 (u+1)^3}{1048576}+
\frac{8085 (u+1)^4}{33554432}+\dots \\
a_2(u)&\sim& \frac{17}{262144} + \frac{721 (u+1)}{8388608}+ \frac{10941(u+1)^2}{134217728}+ \frac{141757 (u+1)^3}{2147483648}
+\frac{3342339 (u+1)^4}{68719476736}
+\dots
\end{eqnarray}
Inserting these expansions into the all-orders Bohr-Sommerfeld expression (\ref{eq:bs2}) we obtain:
\begin{eqnarray}
N+\frac{1}{2}&=&\frac{2}{ \hbar}\left(a_0(u)+\hbar^2 a_1(u) +\hbar^4 a_2(u)+\dots\right) \nonumber \\
&\sim& 
\frac{1}{\hbar}\left((u+1)+\frac{ (u+1)^2}{16} +\frac{3  (u+1)^3 }{256}
+\frac{25 (u+1)^4}{8192}+
 \frac{245 (u+1)^5}{262144} +\dots\right) \nonumber \\
&& +2 \hbar\left( \frac{1}{128}+\frac{5(u+1)}{2048}+ \frac{35 (u+1)^2}{32768} +
\frac{525 (u+1)^3}{1048576}+
\frac{8085 (u+1)^4}{33554432}+\dots \right)  \nonumber \\
&&+2\hbar^3\left(\frac{17}{262144} + \frac{721 (u+1)}{8388608}+ \frac{10941(u+1)^2}{134217728}+ \frac{141757 (u+1)^3}{2147483648}
+\frac{3342339 (u+1)^4}{68719476736}
+\dots\right)  \nonumber \\
&&+\dots
\label{eq:wkb-weak}
\end{eqnarray}
This expresses the band label $N$ as a formal series in $\hbar$, with coefficients that are functions of the energy $u$.
To compare with perturbation theory, we invert to write the energy $u$ as a formal series in $\hbar$, with coefficients that are functions of the band label $N$. Doing so, we obtain precisely the perturbative expansion (\ref{eq:weak}). Notice, that in this latter formal series, the coefficients of the formal series in $\hbar$ are simple polynomials in $\left(N+\frac{1}{2}\right)$. Also recall that this perturbative expansion (\ref{eq:weak}) is non-Borel-summable in this regime where $ N \hbar \ll 1$.

\subsubsection{Band Width at Weak Coupling}

The exponentially narrow band widths can be evaluated using the dual actions. In the weak-coupling region, 
\begin{eqnarray}
u&\sim & -1+2\, a_0(u) +\dots = -1+\hbar\left(N+\frac{1}{2}\right) +\dots \\
\pi \,{\rm Im} [ a_0^D] &\sim& 4+\frac{1+u}{2}\left(\ln\left(\frac{1+u}{32}\right) -1\right)+\dots 
\label{eq:dyonic-limits}
\end{eqnarray}
Therefore, from (\ref{eq:band}) we obtain the band width estimate (using Stirling's formula in the last step):
\begin{eqnarray}
\Delta u_{\rm band} &\sim& \frac{2\hbar}{\pi} \left(\frac{\hbar\left(N+\frac{1}{2}\right)}{32\, e}\right)^{-(N+\frac{1}{2})} \, e^{-8/\hbar} \nonumber\\
&\sim& \sqrt{\frac{2}{\pi}}\frac{2^{4(N+1)}}{N!} \left(\frac{2}{\hbar}\right)^{N-\frac{1}{2}}\, e^{-8/\hbar}
\label{eq:dyonic-band-width}
\end{eqnarray}
in agreement with (\ref{eq:band-width}).

\subsection{Strong Coupling}
\label{sec:wkb-strong}

\subsubsection{Gap Center Location from Bohr-Sommerfeld at  Strong Coupling}

In the strong coupling region (i.e. $u\gg 1$, or $N \hbar\gg 1$) the all-orders Bohr-Sommerfeld condition (\ref{eq:bs}) that identifies $u$ with the center of the gap reads 
\begin{eqnarray}
{\hbar\over2} N = a(u, \hbar)\equiv \sum_{n=0}^\infty \hbar^{2n} a_n(u)
\label{eq:bs3}
\end{eqnarray}
The WKB actions $a_n(u)$ can be expanded for $u\gg 1$ as:
\begin{eqnarray}
a_0(u)&\sim& \sqrt{2u}\left( 1-\frac{1}{16 u^2}-\frac{15}{1024 u^4} -\frac{105}{16384 u^6}-\dots 
\right) \\
a_1(u)&\sim&-\frac{1}{16 \left(2u\right)^{5/2}} \left(1+\frac{35}{32 u^2}+\frac{1155}{1024 u^4}
+\frac{75075}{65536 u^6}+\dots\right)  \\
a_2(u)&\sim&-\frac{1}{64 \left(2u\right)^{7/2}}\left(1+ \frac{273}{64 u^2} +\frac{5005}{512 u^4}+
\frac{2297295}{131072 u^6}+\dots\right)  
\end{eqnarray}
Combining these expansions we find
\begin{eqnarray}
N&=& \frac{2}{ \hbar}\left(a_0(u)+\hbar^2 a_1(u) +\hbar^4 a_2(u)+\dots\right) \nonumber\\
&\sim& \frac{2\sqrt{2 u}}{\hbar} \left[1-\frac{1}{16 u^2}-\frac{15}{1024 u^4}-\frac{105}{16384 u^6} 
-\frac{15015}{4194304 u^8}
-\dots\right] \nonumber\\
&& -\frac{\hbar}{8 (2 u)^{5/2}}\left[ 1+\frac{35}{32 u^2}+\frac{1155}{1024 u^4}
+\frac{75075}{65536 u^6}
+\frac{4849845}{4194304 u^8}
+\dots \right]  \nonumber \\
&&- \frac{\hbar^3}{32 (2 u)^{7/2}} \left[1+ \frac{273}{64 u^2} +\frac{5005}{512 u^4}+
\frac{2297295}{131072 u^6}
+ \frac{115426311}{4194304 u^8}
+ \dots\right] \nonumber\\
&& -\dots 
\label{eq:wkb-strong}
\end{eqnarray}
This expresses the band label $N$ as a formal series in $\hbar$, with coefficients that are functions of the energy $u$.
To compare with the strong-coupling expansion, we invert to write the energy $u$ as a formal series in $\hbar$, with coefficients that are functions of the band label $N$. This leads to (here we have defined the ``action''
$
a\equiv \frac{N\, \hbar}{2}$):
\begin{eqnarray}
u&\sim& \left[{a^2\over2}+ \frac{1}{4\,a^2}+\frac{5}{64} \frac{1}{a^6}+\frac{9}{128} \frac{1}{a^{10}}+\dots\right]
+\hbar^2\left[\frac{1}{16\,a^4}+\frac{21}{128} \frac{1}{a^8}+\frac{55}{128} \frac{1}{a^{12}}+\dots \right]
+\dots \\
&\sim& {1\over 2}\left(\frac{N\hbar}{2}\right)^2+ \frac{1}{4} \left(\frac{2}{N\hbar}\right)^2 \frac{1}{\left(1-\frac{\hbar^2}{(N \hbar)^2}\right)}+\frac{5}{64} \left(\frac{2}{N\hbar}\right)^6\frac{\left(1+\frac{7\hbar^2}{5(N \hbar)^2}\right)}{\left(1-\frac{\hbar^2}{(N \hbar)^2}\right)^3\left(1-\frac{4\hbar^2}{(N \hbar)^2}\right)}+\dots 
\label{eq:smallq-u}
\end{eqnarray}
 We recognize precisely the strong-coupling expansion (\ref{eq:strong}). 
Thus, the all-orders-WKB action $a(u, \hbar)$ determines the (convergent) expansion of the location of the gap high up in the spectrum.

\subsubsection{Gap Width at Strong Coupling}

Despite these expressions (for the center of the $N^{\rm th}$ gap) being convergent, there are still non-perturbatively small corrections associated with the narrow gaps high in the spectrum (see Figure \ref{fig:fig1}). Expanding the leading actions in this spectral region we find
\begin{eqnarray}
u&\sim & \frac{1}{2} a_0^2 +\dots  = \frac{\hbar^2}{8} N^2 +\dots \\
\pi \,\mc Im[ a_0^D] &\sim& \sqrt{2u}\left(\ln(8 u)-2\right)+\dots 
\label{eq:electric-limits}
\end{eqnarray}
Therefore, from (\ref{eq:gap}) we obtain the gap width estimate:
\begin{eqnarray}
\Delta u_{\rm gap} &\sim& \frac{\hbar^2 N}{2\pi} \left(\frac{e}{\hbar N}\right)^{2N} 
\label{eq:electric-gap-width}
\end{eqnarray}
in agreement with (\ref{eq:gap-width}). Thus, the formulas (\ref{eq:band}, \ref{eq:gap}) have the correct form in both extreme limits, in one case referring to the width of a band, and in the other to the width of a gap.

\subsection{Intermediate Coupling: Instanton Condensation Near the Barrier Top}
\label{sec:top}

\begin{figure}[htb]
\includegraphics[scale=0.7]{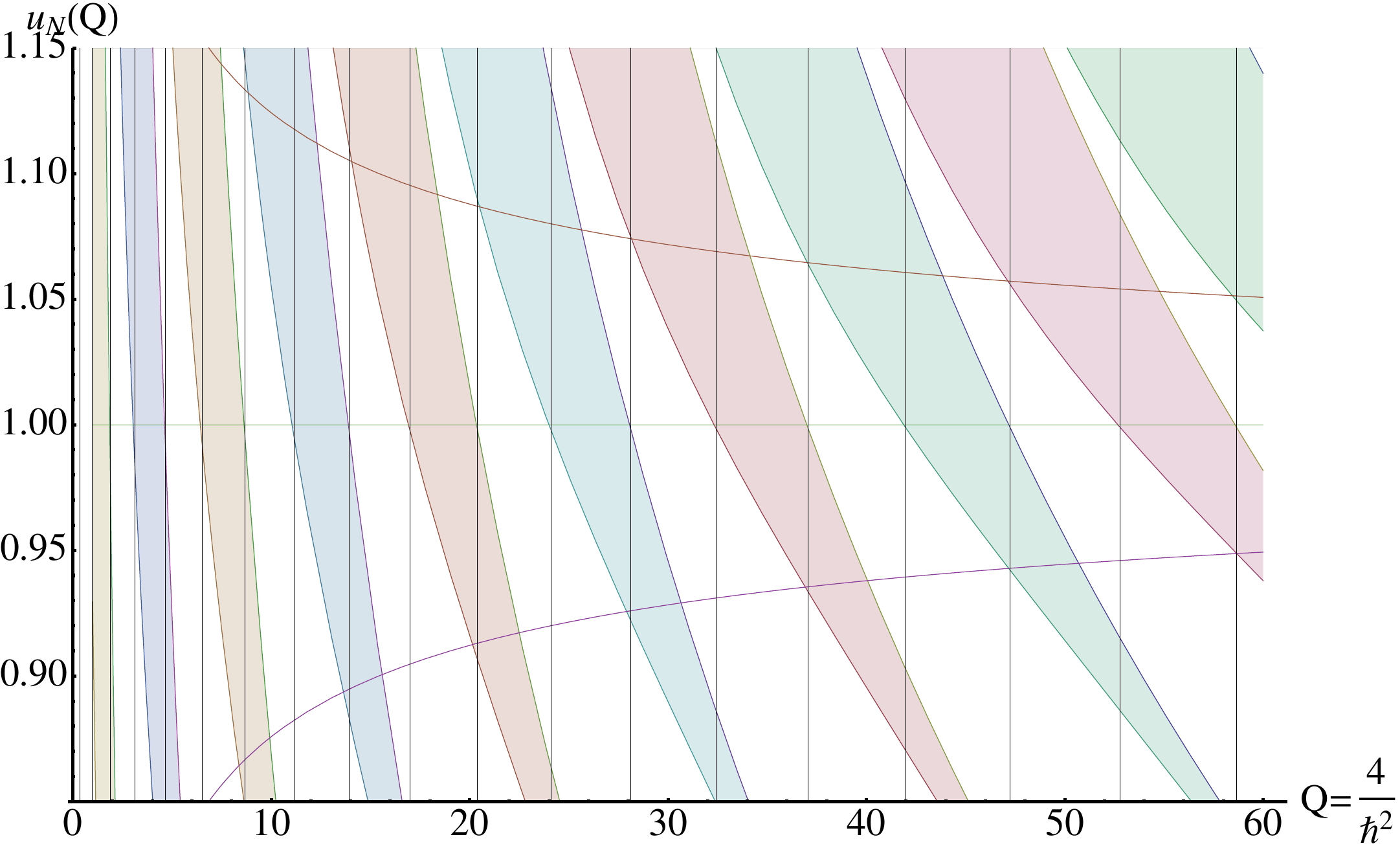}
\caption{The bands (shaded) and gaps,  as a function of the parameter $Q\equiv\frac{4}{\hbar^2}$, near  the barrier top  $u=1$. The vertical lines denote the values $Q=\frac{\pi^2}{16}\left(N\pm \frac{1}{4}\right)^2$, where $N$ is the band label, and which agree accurately with the points at which the band/gap edges intersect the line $u=1$. The curved solid lines show the expressions in (\ref{eq:top-gaps}), which gives the gap splitting at the top of a band, centered at $u=1$. This leading expression fits the exact curves quite well.}
\label{fig:fig2}
\end{figure}

Near the barrier top, where $u\sim 1$, there is a transition from the divergent perturbative behavior characteristic of the weak coupling region, to the convergent perturbative expansions characteristic of the strong-coupling region. As is clear from the plots in Figures \ref{fig:fig1} and \ref{fig:fig2}, the transition is smooth, but connecting the regions requires careful interpretation of the various expansions. Of particular interest are the different mechanisms by which non-perturbative terms arise in the different physical regions. For example, the general expressions for the exponentially narrow width of a band (\ref{eq:band}) deep in the weak coupling region, and of a gap (\ref{eq:gap}) high in the strong coupling region, are not valid in the region in the vicinity of the barrier top, because the single-instanton factor $\exp\left[-\frac{2\pi}{\hbar}\, {\rm Im}\, a_0^D\right]$  is no longer exponentially small. Thus, the explicit expressions  (\ref{eq:band-width}) and  (\ref{eq:gap-width}) are not accurate in this region of the spectrum.  Physically, this is  a region  of ``instanton condensation''. This is of particular interest as it is a direct analogue of an instanton condensation phenomenon observed in matrix models and 2d gauge theories \cite{gw,wadia,neuberger,matytsin,witten}.

In the spectral region near $u\sim 1$:
\begin{eqnarray}
a_0&\sim& {4\over \pi} +\frac{u-1}{2\pi}\left[\ln\left(\frac{32}{u-1}\right) +1\right]+\dots \\
-i a_0^D&\sim& \frac{1}{2}\left(u-1\right)+\dots
\label{eq:barrier-limits}
\end{eqnarray}
The fact that $a_0^D$ vanishes at $u=1$ implies that the single-instanton approximations (\ref{eq:band-width}, \ref{eq:gap-width}) are no longer good, as the exponentially small instanton factor $\exp\left[-\frac{2\pi}{\hbar}\, {\rm Im}\, a_0^D\right]$  is now of order 1. The fact that $a_0(u)$ tends to a non-zero constant implies the leading scaling between $N$ and $\hbar$ in this region:
\begin{eqnarray}
N+\frac{1}{2}&\sim& \frac{8}{\pi\,\hbar}\qquad (\text{band center at } u=1)
\label{eq:topb} \\
N&\sim& \frac{8}{\pi\,\hbar}\qquad (\text{gap center at } u=1) 
\label{eq:topg}
\end{eqnarray}
We had already expected that the barrier top would be in the vicinity of  $N \sim 1/\hbar$, but the behavior  $a_0(u)\sim 4/\pi$ fixes the non-trivial coefficient to be $8/\pi$. We can now use these estimates (\ref{eq:topb}, \ref{eq:topg})  in either  the weak-coupling expansion (\ref{eq:weak}) or the strong coupling expansion (\ref{eq:strong}) to obtain two very different looking expressions for the energy at the top of the barrier:
\begin{eqnarray}
u_{\rm weak}&\sim& -1+\frac{8}{\pi}\left[1-\frac{1}{16}\frac{8}{\pi}-\frac{1}{2^8}\left(\frac{8}{\pi}\right)^2-\frac{5}{2^{14}}\left(\frac{8}{\pi}\right)^3 -\frac{33}{2^{18}}\left(\frac{8}{\pi}\right)^4-\dots \right] +O(\hbar)\nonumber\\
&=& 1+O(\hbar)  \\
u_{\rm strong}&\sim& \frac{1}{2} \left[\left(\frac{4}{\pi}\right)^2 + \frac{1}{2} \left(\frac{\pi}{4}\right)^2+\frac{5}{32} \left(\frac{\pi}{4}\right)^6+\frac{9}{64}\left(\frac{\pi}{4}\right)^{10}+\dots \right]+O(\hbar) \nonumber\\
&=&1+O(\hbar)
\label{eq:barriertop}
\end{eqnarray}
It is interesting that these two  different expansions agree  at $u\sim 1$, but there are corrections  $\sim O(\hbar)$, corresponding to the band and gap widths in the vicinity of the barrier top.

The estimates in (\ref{eq:topb}, \ref{eq:topg})   can be refined further:   the {\it edges} of the bands/gaps when $u=1$ are given by (see also \cite{keller}):
\begin{eqnarray}
N\pm \frac{1}{4}\sim \frac{8}{\pi \hbar} \qquad (\text{band/gap  edge at } u=1)
\label{eq:refined-top}
\end{eqnarray}
as shown in Figure \ref{fig:fig2}. In this Figure, showing the bands (shaded) and gaps (unshaded), plotted as a function of the parameter $Q=\frac{4}{\hbar^2}$ that appears in the Mathieu equation (\ref{mathieu1}, \ref{hbar}), we see clearly that at one of these edges, the band and the gap immediately above and below have equal width. Physically, this is directly related to the fact that the discriminant can be expressed in terms of the reflection and transmission amplitudes \cite{keller-scattering}, and the fact that at the top of an inverted harmonic barrier the reflection and transmission probabilities are equal, both being 1/2 \cite{ford}. Relatively little is known rigorously about the corrections to this behavior of the band/gap widths in this region, even though in some sense it is the most interesting region physically \cite{keller,connor,Basar:2015xna}. The results of Weinstein and Keller imply that the small $\hbar$ behavior for the upper band edges has leading behavior
\begin{eqnarray}
u\sim 1\pm \frac{\pi \hbar}{16}+\dots 
\label{eq:top-gaps}
\end{eqnarray}
These lines are shown in Figure \ref{fig:fig2}, and we see that they do indeed intersect the upper band edges quite accurately. It would be interesting to investigate more precisely  the band and gap widths in this region.

\subsection{Summary of Different Behaviors of Mathieu Spectrum in Different Regions}
\label{sec:summary}

It is worth comparing the form of the expressions for the energy in the three different spectral regions.
\\

$\bullet$ 
{\bf Weak coupling:} $N \hbar\ll 1$, deep inside the wells, far below the barrier top. Here the band edge  energy eigenvalues $u^{(\pm)}(\hbar, N)$ are expressed as a resurgent trans-series:
\begin{eqnarray}
u^{(\pm)}(\hbar, N)\sim\sum_{n=0}^\infty u_n(N)\hbar^n \pm 
\frac{32}{\sqrt{\pi} \, N!}\left(\frac{32}{\hbar}\right)^{N-1/2}
\exp\left[-\frac{8}{\hbar}\right] \sum_{n=0}^\infty d_n(N)\hbar^n 
+\dots
\label{eq:bw}
\end{eqnarray}
The first term is the formal perturbative series, $u_{\rm pert}(\hbar, N)$, which is divergent and Borel non-summable. The leading Borel poles occur at the two-instanton location. The first, one-instanton,  exponential correction in (\ref{eq:bw}) gives the leading band width.
Furthermore, the expansion coefficients $u_n(N)$ and $d_n(N)$ are explicitly related: the $d_n(N)$ are fully determined by the $u_n(N)$, as in (\ref{eq:one}, \ref{eq:prefactor}).
\\

$\bullet$
{\bf Strong coupling:} $N \hbar\gg1$, far above the barrier top. Here the gap edges are conventionally written as the convergent expansions (\ref{eq:conv}). This raises an obvious question: how can such convergent expansions smoothly connect to the trans-series expansion (\ref{eq:bw}) below the barrier top? In fact \cite{Basar:2015xna}, one can re-write the expansions (\ref{eq:conv}) in the following suggestive form:
\begin{eqnarray}
\hskip -.75cm  u^{(\pm)}(\hbar, N)=\frac{\hbar^2  N^2}{8}\sum_{n=0}^{N-1}  \frac{\alpha_n(N)}{\hbar^{4n}} \pm \frac{\hbar^2}{8}{\left(\frac{2}{\hbar}\right)^{2N}\over (2^{N-1}(N-1)!)^2}  \sum_{n=0}^{N-1}\frac{\beta_n(N)}{\hbar^{4n}} +\dots
\label{eq:smallq-alt}
\end{eqnarray}
where the coefficient functions $\alpha_n(N)$ and $\beta_n(N)$ are rational functions of the gap label $N$, but have poles at the two-instanton location. Thus the expansions  truncate at $n=N$. Note that this has a similar form to the trans-series in (\ref{eq:bw}). As $N\to\infty$, the polynomial expansions  in $\frac{1}{\hbar^4}$ extend to infinite series. Remarkably, with this re-arrangement we find once again that the expansion coefficients $\alpha_n(N)$ and  $\beta_n(N)$  are explicitly related: the  $\beta_n(N)$  are fully determined by the $\alpha_n(N)$. Thus, the strong-coupling expansions can also be written in a form that matches the trans-series structure of the weak-coupling region \cite{bdu2}.
\\

$\bullet$ 
{\bf Transition region: intermediate coupling:} $N \hbar \sim \frac{8}{\pi}$, near  the barrier top. This is the instanton condensation region, where the single-instanton exponential factor becomes of order 1, so that all instanton orders need to be taken into account. Here, relatively little is known precisely. One concrete statement is that the bands and gaps in this region have equal width at leading order, and this width is $O(\hbar)$, for example as in (\ref{eq:top-gaps}).

\section{Path Integral Interpretation: Steepest Descents and Lefschetz Thimbles}
\label{sec:path}
 
We have used WKB techniques in the framework of differential equations to derive the new result (\ref{eq:one}, \ref{eq:prefactor}), which relates higher orders of the non-perturbative trans-series expansion to the perturbative series. But the result is even more interesting when viewed from the equivalent path integral language. Here, different instanton sectors correspond to different saddle points of the path integral, and so we learn that there is a saddle point expansion of the path integral representation of the quantum resolvent, in which the fluctuations about different saddle points are directly related to one another. In practice, this has been confirmed by the work of \cite{Escobar-Ruiz:2015rfa} because their diagrammatic computation of the fluctuations about the one-instanton sector is based on a perturbative expansion of the path integral about the one-instanton sector. But the fact that this computation is so technically difficult means that there should be a simpler way to understand the final result. We propose that this is an example where a saddle point expansion in terms of Lefschetz thimbles can  be implemented explicitly.

To illustrate the basic idea, we turn back to a zero-dimensional example, and recall the results of Berry and Howls \cite{BerryHowls} concerning an all-orders steepest descents expansion of an ordinary contour integral. They considered a contour integral of the form
\begin{eqnarray}
I^{(n)}(\hbar)=\int_{C_n}dz\, e^{-\frac{1}{\hbar}\, f(z)}={\color{red} \frac{1}{\sqrt{1/\hbar}}\, \exp\left[-\frac{1}{\hbar}\, f_n \right]}\, {\color{blue}T^{(n)}(\hbar)}
\label{eq:bh1}
\end{eqnarray}
where the contour $C_n$ follows the steepest descent path through the saddle point $z_n$ of the function $f(z)$. The first two factors on the RHS are the dominant exponential term and the  Gaussian fluctuation prefactor. The remaining factor, $T^{(n)}(\hbar)$, is a formal series in $\hbar$, and represents all orders of the perturbative fluctuations about this saddle point. Physicists typically just keep the exponential and the Gaussian prefactor. Berry and Howls \cite{BerryHowls} point out that there is a great deal of interesting information encoded in the further fluctuations $T^{(n)}(\hbar)$. For example, under relatively mild assumptions about the function $f(z)$, one can show by straightforward contour deformation that the fluctuations about different saddle points are directly related. This is a manifestation of Darboux's theorem: the fluctuations about a given point are governed by the nearest singularity, which here is the nearest (connected) saddle point \cite{BerryHowls}. This fact can be expressed as the following integral transform:
\begin{eqnarray}
T^{(n)}(\hbar)=\frac{1}{2\pi\, i} \sum_m (-1)^{\gamma_{nm}} \int_0^\infty \frac{dv}{v} \frac{e^{-v}}{1-\hbar  v/(F_{nm})}\, T^{(m)}\left(\frac{F_{nm}}{v}\right)
\label{eq:bh2}
\end{eqnarray}
where the sum is over all other saddles topologically connected to the $n^{\rm th}$ saddle, and the $\gamma_{nm}$ determine the orientations of the deformed steepest descent contours. Here $F_{nm}\equiv f_m-f_n$ is the difference of the exponents at the saddle points $z_m$ and $z_n$. An immediate consequence of this result  is that if we write the fluctuation about  the $n^{\rm th}$ saddle as
\begin{eqnarray}
T^{(n)}(\hbar)\sim \sum_{r=0}^\infty T^{(n)}_r\, \hbar^{r}
\label{eq:bh3}
\end{eqnarray}
then expanding both sides of (\ref{eq:bh2}) we find a relation between the fluctuation coefficients around saddle points $z_n$ and $z_m$:
\begin{eqnarray}
T^{(n)}_r \sim  \frac{(r-1)!}{2\pi\, i} \sum_m  \frac{(-1)^{\gamma_{nm}}}{\left(F_{nm}\right)^r}\left[T^{(m)}_0+\frac{F_{nm}}{(r-1)}\, T^{(m)}_1+\frac{\left(F_{nm}\right)^2}{(r-1)(r-2)}\, T^{(m)}_2+\dots \right]\quad, \quad r\to\infty
\label{eq:bh4}
\end{eqnarray}
Notice the  universal large-order factorial divergence of fluctuations, and more importantly notice that this says that fluctuations about {\it neighboring} saddles are explicitly related. 

For full details see \cite{BerryHowls}, but here it is more appropriate to illustrate how this works with some examples. Consider the zero-dimensional ``partition function'' for the Mathieu system, writing the periodic potential as $V(z)=\sin^2(z)$:
\begin{eqnarray}
Z(\hbar)=\int_{-\pi/2}^{\pi/2} dz\, e^{-\frac{1}{\hbar}\, \sin^2(z)}
\label{eq:0dz}
\end{eqnarray}
There are two saddle points: $z_0=0$ and $z_1=\frac{\pi}{2}$. 
It is straightforward to generate the ``perturbative expansion'' about the ``vacuum'' saddle point at $z_0=0$. The expansion coefficients in (\ref{eq:bh3}) for $n=0$ are:
\begin{eqnarray}
T^{(0)}_r  &=& \frac{\Gamma\left(r+\frac{1}{2}\right)^2}{\sqrt{\pi}\, \Gamma(r+1)} \nonumber \\
& \sim&  \frac{(r-1)!}{\sqrt{\pi}}\left({\color{blue}1}-\frac{{\color{blue}\frac{1}{4}}}{(r-1)}+\frac{{\color{blue}\frac{9}{32}}}{(r-1)(r-2)}-\frac{{\color{blue}\frac{75}{128}}}{(r-1)(r-2)(r-3)}
+\dots\right) \quad, \quad r\to\infty
\label{eq:berry-d00c}
\end{eqnarray}
It is also straightforward to generate the  fluctuation expansion about the ``non-perturbative saddle'' at $z_1=\frac{\pi}{2}$:
\begin{eqnarray}
T^{(1)}(\hbar)\sim i\, \sqrt{\pi} \left({\color{blue}1}-{\color{blue}\frac{1}{4}}\, g^2
+{\color{blue}\frac{9}{32}}\, g^4-{\color{blue}\frac{75}{128}}\, g^6+\dots \right)
\label{eq:berry-d00d}
\end{eqnarray}
Comparing coefficients, we see that the low-order coefficients of the fluctuations about the non-perturbative saddle $z_1$ govern the large-order behavior of the fluctuations about the vacuum saddle $z_0$. Below we consider a more interesting example where there are three saddles.

This is a general feature of a wide class of ordinary exponential integrals: fluctuations around different saddles are quantitatively related. In studying path integrals, and especially semiclassical  multi-instanton expansions of path integrals, we often motivate our formal manipulations by analogy with steepest descents asymptotic analysis of ordinary integrals. Thus, it is a reasonable question to ask whether or not something like this could possibly occur for (infinite dimensional!) path integrals. This leads to the theory of infinite dimensional Morse Theory and Lefschetz thimbles. Even the generalization from a one-dimensional integral to a multi-dimensional integral introduces interesting and highly non-trivial effects \cite{Fedoryuk,arnold,Pham,delabaere2002}. There are many more subtleties in going from finite to infinite dimensions \cite{Witten:2010zr,Guralnik:2007rx,Kontsevich-1}.

Instead of delving into this unresolved issue, let us illustrate what happens when you ask this question about the Mathieu system. Here we have done the calculation without using the path integral, but we can re-interpret the results in path integral terms. Specifically, consider the Schr\"odinger problem with periodic potential $V(x)=\sin^2(x)$. The  large order growth of the perturbative expansion coefficients for the ground state energy is \cite{Stone:1977au}
\begin{eqnarray}
c_n\sim n!\left({\color{blue}1}-{\color{blue}\frac{5}{2}}\cdot\frac{1}{n}-{\color{blue}\frac{13}{8}}\cdot\frac{1}{n(n-1)}-\dots\right)
\label{eq:berry-d1sga}
\end{eqnarray}
This is the large-order growth of the fluctuation about the
 vacuum saddle point, for the ground state energy. Next, we can inspect the multi-instanton trans-series expansion for the same physical quantity, the ground state energy, and look for the fluctuations about the `nearest' saddle point with the same quantum numbers as the perturbative vacuum. This is the instanton/anti-instanton saddle point, and we find  \cite{Dunne:2013ada}
 \begin{eqnarray}
{\rm Im}\, E_0 \sim \pi\, e^{-2 \frac{1}{2 \hbar}}\left({\color{blue}1}-{\color{blue}\frac{5}{2}} \cdot\hbar-
{\color{blue}\frac{13}{8}} \cdot \hbar^2-\dots\right)
\label{eq:berry-d1sgb}
\end{eqnarray}
Notice the correspondence of the factors appearing in these different expansions, about different saddles.
This is very surprising. It means that to some degree the basic resurgent relation between large-order behavior of fluctuations about one saddle and low-orders of fluctuations about a neighboring saddle is indeed inherited by the path integral. (This is not just a feature of the Mathieu system: a similar relation holds also for the double-well potential).

To emphasize that this structure could be quite general we consider a further example that has the interesting feature of having more saddle points. Let us generalize the zero-dimensional example above, by replacing the periodic Mathieu potential $V(x)=\sin^2(x)$ by the doubly-periodic elliptic potential of Lam\'e type \cite{Basar:2013eka}: 
\begin{eqnarray}
V(x)={\rm sd}^2(x| m)\qquad, \qquad 0\leq m\leq 1
\label{eq:lame}
\end{eqnarray}
where $0\leq m\leq 1$ is the elliptic parameter. Note that this potential interpolates smoothly between the periodic Mathieu case, $V(x)=\sin^2(x)$, when $m=0$ and the ``Sinh-Gordon'' potential, $V(x)=\sinh^2(x)$, when $m=1$. We write the  zero dimensional partition function as  the trace over the period:
\begin{eqnarray}
Z(\hbar | m)=\frac{1}{\sqrt{\hbar \pi}} \int_{-\mathbb K}^{\mathbb K}dz\, e^{-{1\over\hbar} \,{\rm sd}^2(z | m)}
\label{eq:0dlame}
\end{eqnarray}
This has the following ``perturbative expansion'' about the ``vacuum saddle'' at $z=0$.
\begin{eqnarray}
Z_{pert}(\hbar  | m)=\sum_{n=0}^\infty a^{(0)}_n(m)\,\hbar^{n}
\label{eq:0dl1}
\end{eqnarray}
For example, at some chosen values of $m$ we have: 
\begin{eqnarray}
 Z_{\rm pert}\left(\hbar | 0\right) &=&1+ \frac{1}{4} \hbar+ \frac{9}{32}  \hbar^2 + \frac{75}{128}  \hbar^3 + \frac{3675 }{2048}\hbar^4 + \frac{59535}{8192}\hbar^5 + \ldots \nonumber\\
Z_{\rm pert}\left(\hbar | 1\right) &=&1- \frac{1}{4} \hbar + \frac{9}{32} \hbar^2 - \frac{75}{128} \hbar^3 + \frac{3675}{2048} \hbar^4 -\frac{59535}{8192} \hbar^5 
+ \ldots \nonumber\\
Z_{\rm pert}\left(\hbar \bigg | \frac{1}{4} \right) &=&  1+ \frac{1}{8} \hbar + \frac{9}{64} \hbar^2 + \frac{105}{512} \hbar^3 + \frac{1995}{4096} \hbar^4 + \frac{48195}{32768} \hbar^5 + \dots \nonumber\\
Z_{\rm pert}\left(\hbar \bigg | \frac{3}{4} \right) &=&  1- \frac{1}{8} \hbar + \frac{9}{64} \hbar^2 -\frac{105}{512} \hbar^3 + \frac{1995}{4096} \hbar^4 - \frac{48195}{32768} \hbar^5 + \dots \nonumber\\
Z_{\rm pert}\left(\hbar \bigg | \frac{1}{2} \right) &=&1+   0\, \hbar + \frac{3}{32} \hbar^2 +  0\, \hbar^3 + \frac{315}{2048} \hbar^4 +0\,\hbar^5 +  \ldots 
\label{eq:0de}
\end{eqnarray} 
These perturbative expansions are divergent for all $m$, but are non-alternating for $m<1/2$, and alternating for $m>1/2$. This latter fact reflects  the duality relation: $Z(\hbar | m)= Z(-\hbar | 1-m)$, which follows from a property of the Jacobi elliptic function sd.

The ``action''  function $V(z)$ has two different types of non-trivial saddles, one set along the real axis and another on the imaginary axis. Their relative distance from the vacuum saddle at $z=0$ is governed by the value of the elliptic parameter $m$. At the real saddle point $z_1=\mathbb K(m)$, we have the action $S_1=\frac{1}{1-m}$, while at the imaginary saddle at $z_2=i \mathbb K(1-m)$ we have $S_2=-\frac{1}{m}$. These two different saddles can be seen in the large-order behavior of the perturbative fluctuation coefficients. Numerically, by studying the large-order behavior of the expansion coeffiicents about $z_0=0$,  one finds \cite{Basar:2013eka}
\begin{eqnarray}
a^{(0)}_n(m)\sim\frac{(n-1)!}{\pi}\left( S_1^{n+1/2}(m)+(-1)^n |S_2|^{n+1/2}(m)\right)
\label{eq:0dl2}
\end{eqnarray}
For $m<\frac{1}{2}$, $S_1$ dominates and the $a^{(0)}_n(m)$ are non-alternating in sign. For $m>\frac{1}{2}$, $S_2$ dominates and the $a^{(0)}_n(m)$ alternate in sign. For $m=\frac{1}{2}$, $S_1$ and $S_2$ are equal in magnitude, and the odd terms vanish due to interference cancellations. 
Thus both the real and complex saddles influence the large-order growth of the fluctuations about the trivial vacuum saddle at $z_0=0$. This is an explicit example of the sum of neighboring saddles in (\ref{eq:bh4}).

Actually, this relation can be made even more precise \cite{Basar:2013eka}, noting that the fluctuations about the saddles $z_1=\mathbb K(m)$ and $z_2=i \mathbb K(1-m)$ can be written as 
\begin{eqnarray}
Z^{(1)}\left(\hbar | m\right) &=& i\, e^{-S_1/\hbar}\sum_{n=0}^\infty a_n^{(1)}(m) \hbar^n \\
Z^{(2)}\left(\hbar | m\right) &=& i\, e^{-S_2/\hbar}\sum_{n=0}^\infty a_n^{(2)}(m) \hbar^n
\label{eq:0dl4}
\end{eqnarray}
and one finds an exact  resurgence relation:
\begin{eqnarray}
Z^{(0)}\left(\hbar | m\right)={2\over 2\pi i}\hskip -.1cm \sum_{k\in \{1,2 \}} \int_0^\infty {dv\over v}{1\over1-\hbar\,v} Z^{(k)}(v |m)
\label{eq:0dlameresurgence}
\end{eqnarray}
and an exact relation between the expansion coefficients $a_n^{(k)}$:
\begin{eqnarray}
a^{(0)}_n(m) =\sum_{j=0} \frac{(n-j-1)!}{\pi}\,
\left({a^{(1)}_j(m)\over S_1^{\,n-j}}+{a^{(2)}_j(m)\over S_2^{\,n-j}}\right)
\label{eq:0dl5}
\end{eqnarray}

Now we can ask again how much of this resurgent structure is inherited by the {\it path integral}  version of this problem. The partition function is now an {\it infinite dimensional functional integral}:
\begin{eqnarray}
Z(\hbar |m)=\int {\mathcal D} x\, e^{- S[x]}  = \int {\mathcal D}x\, e^{-\int d\tau \left( \frac{1}{4}  
\dot x^2 + \frac{1}{\hbar}\, {\rm sd}^2(\sqrt{\hbar}\, x | m) \right)} 
\label{eq:1dlamez}
\end{eqnarray}
As before for the Mathieu case, we illustrate our point using the simplest observable, the perturbative ground state energy, for which we find perturbative expansions as functions of $m$. For several selected values of $m$ the first few terms are:
  \begin{eqnarray}
E^{(0)}\left(\hbar | 0 \right)&=&1-\frac{1}{4} \hbar -\frac{1}{16} \hbar^2 -\frac{3}{64} \hbar^3 -\frac{53 }{1024} \hbar^4 -\frac{297}{4096}\hbar^5  -\dots 
\nonumber\\
E^{(0)}(\hbar | 1)&=& 1+\frac{1}{4} \hbar -\frac{1}{16} \hbar^2 +\frac{3 }{64} \hbar^3 -\frac{53}{1024} \hbar^4 +\frac{297}{4096} \hbar^5 -\dots 
\nonumber\\
E^{(0)}\left(\hbar \bigg | \frac{1}{4}\right) &=&1-\frac{1}{8} \hbar -\frac{11}{128} \hbar^2 -\frac{3}{128} \hbar^3 -\frac{889}{32768} \hbar^4 -\frac{225}{8192} \hbar^5 -\dots 
\nonumber\\
E^{(0)}\left(\hbar \bigg | \frac{3}{4}\right) &=&1+\frac{1}{8} \hbar -\frac{11}{128} \hbar^2 +\frac{3}{128} \hbar^3 -\frac{889}{32768} \hbar^4 +\frac{225}{8192} \hbar^5 -\dots 
\nonumber\\
E^{(0)}\left(\hbar \bigg | \frac{1}{2}\right)&=&1+0\, \hbar -\frac{3}{32} \hbar^2 +0 \, \hbar^3 -\frac{39}{2048} \hbar^4 +0\, \hbar^5
-\dots
\label{eq:1dlame1}
\end{eqnarray}
These expansions are remarkably similar to the zero dimensional expansions of the zero dimensional partition function  in (\ref{eq:0de}). For example, we again observe the duality relation: $E^{(0)}(\hbar | m)= E^{(0)}(-\hbar | 1-m)$, and the perturbative expansions are non-alternating for $m<\frac{1}{2}$, but alternating for $m>\frac{1}{2}$.

The Lam\'e  potential is doubly periodic, in the complex plane. This means that there are two types of instantons: real instantons and complex ``ghost'' instantons. The real instantons tunnel between minima of the potential along the real axis, while the complex instantons tunnel between saddles along the imaginary axis. The associated classical actions are:
\begin{eqnarray}
S_{\cal I}(m) &=&\frac{2\arcsin(\sqrt{m})}{\sqrt{m(1-m)}}\\
 S_{\cal G}(m) &=&\frac{-2\arcsin(\sqrt{1-m})}{\sqrt{m(1-m)}}
 \label{eq:realcomplex} 
\end{eqnarray}
Then one can study the large order growth of the perturbative expansion coefficients for the ground state energy and one finds that both the real and complex instantons contribute to this large-order behavior \cite{Basar:2013eka}:
\begin{eqnarray}
a_n(m)\sim -\frac{16}{\pi} n!\left(\frac{1}{( S_{\cal I \bar{\cal I}}(m))^{n+1}}-\frac{(-1)^{n+1}}{| S_{\cal G \bar{\cal G}}(m)|^{n+1}}\right)
\label{eq:both}
\end{eqnarray}
This is remarkable. It means that the complex instantons directly affect perturbation theory, even though they are not in the original path integral measure, which is a sum over all real paths. The close resemblance to the resurgent structure of the zero-dimensional analogue system [see Eq. (\ref{eq:0dl2})] strongly suggests once more that analytic continuation of path integrals may inherit resurgent structure.

Thus we would be tempted to define a path integral by its resurgent thimble expansion:
\begin{eqnarray}
\int {\mathcal D}A\, e^{-\frac{1}{\hbar}S[A]} =\sum_{{\rm thimbles}\, k} {\mathcal N}_k\, e^{-\frac{i}{\hbar}\, S_{\rm imag}[A_k]}  \int_{\Gamma_k} {\mathcal D}A\, 
e^{-\frac{1}{\hbar}S_{\rm real}[A]}
\end{eqnarray}
where the sum over Lefschetz thimbles is the analogue of the sum over steepest descent contours, but now is a sum over  functional  steepest descents contours, or thimbles. This type of expansion is purely formal, but the previous two examples illustrate that there is some truth to it. We take this as motivation to formalize such expansions more rigorously.

\section{Conclusion}

In these lectures we have reviewed a number of different approaches to the general question of resurgent asymptotic expansions for spectral problems, using the Mathieu equation as a concrete example. We have found several interesting new facts, which we believe motivate a more rigorous mathematical analysis of these problems. We hope we have convinced the reader, both physicist and mathematician alike, that there are some interesting new features still lurking in this very old problem, waiting to be made more precise. One of the main conclusions is that there is more structure when one considers the eigenvalue $u$ as a function of two variables, both the level/band/gap label $N$ as well as the coupling $\hbar$. This has surprisingly close analogies with ``large-$N$'' methods in physics \cite{marcos-book}. This also motivates the problem of understanding better the mathematical properties of trans-series of more than one variable. The path integral interpretation of these results, and the close relation to multi-instanton calculus for quantum field theories (both supersymmetric and non-supersymmetric), encourages us to reconsider more seriously the question of defining an all-orders steepest descents expansion of path integrals. Some initial work along these lines has appeared recently \cite{Behtash:2015zha}.

\section{Acknowledgments}

We thank the organizers, in particular Fr\'ed\'eric Fauvet and David Sauzin, for organizing such an interesting workshop in such an inspiring location. We acknowledge support from the US DOE grants  DE-SC0010339 (GD) and DE-SC0013036 (M\"U).
M.\"U.'s work was partially supported by the Center for Mathematical Sciences and Applications  at Harvard University.

\end{document}